\documentclass[showpacs,preprint,prb,aps]{revtex4}
\usepackage{amssymb}
\usepackage{epsfig}
\usepackage{subfigure}
\begin{document}
\title{Damping and decoherence of Fock states in a nanomechanical resonator due to two level systems }

\author{Laura G. Remus and Miles P. Blencowe}\affiliation{Department of Physics and Astronomy, Dartmouth College, Hanover, New Hampshire
03755, USA }

\date{\today}

\begin{abstract}
We numerically investigate the decay of initial  quantum Fock states and their superpositions for a mechanical resonator mode coupled to an environment comprising interacting, damped
tunneling two level system (TLS) defects. The cases of one, three, and six near resonant, interacting TLS's are considered in turn and it is found that the resonator displays Ohmic bath like decay behavior with  as few as three  TLS's. 
\end{abstract}


\maketitle

\section{\label{sec:introduction}Introduction}
The quest to understand the quantum-to-classical transition has led to the development of macroscopic mechanical systems in which researchers hope to realize quantum states.  In a 2010 landmark experiment,\cite{oconnell10} a state corresponding to a single quantum of vibrational energy in a mechanical resonator was created and its subsequent decay dynamics measured. We anticipate that similar measurements involving higher number quantum Fock states in a mechanical system will be achieved in the near future.   In light of these developments, there is a need to understand the decoherence mechanisms in these systems. 

In the early 70's, researchers studying the unusual thermal and acoustic properties of amorphous insulators at cryogenic temperatures suggested that the materials might be populated by tunneling two-level system (TLS) defects.\cite{anderson72, phillips72} More recently, evidence for the relevance of TLS's in  micronscale superconducting qubit dynamics has been established.\cite{constantin09,ku05,martinis05,neeley08,oconnell08,shnirman05,simmonds04,tian07}  The same amorphous materials are often used in the fabrication of micronscale mechanical systems and thus it is likely that TLS's will play a dominant role in their quantum-to-classical transition at cryogenic temperatures.

In particular, we anticipate that TLS's will provide one of the main mechanisms for the decay of quantum states in mechanical resonators.  In Ref.~[\onlinecite{remus09}], we presented an estimate indicating that a given low order flexural mode of a micronscale mechanical resonator vibrating at radio frequencies may be near resonance with a few TLS's, but is unlikely to interact resonantly with large numbers of TLS's.   These TLS's couple to the motion of the resonator via its strain, and thus will be part of the environment responsible for the decay of quantum flexural modes. Reference~[\onlinecite{remus09}] numerically investigated the damping  of initially coherent states and the decoherence dynamics of initial superpositions of spatially separated coherent states, where the environment consisted of either one or three damped TLS's. Clear signatures of resonator amplitude dependence were observed in the damping and decoherence dynamics, a consequence of TLS saturation. This behavior is qualitatively different from the amplitude-independent damping and decoherence resulting from the standard, Ohmic oscillator bath model of an environment. However, it is of interest to explore the damping and decoherence dynamics as the number of near resonant TLS's increases, in particular to establish the expected transition to approximately Ohmic like behavior.    

In this work, we numerically model the low temperature ($\hbar\omega\ll k_B T$) damping and decoherence dynamics of  a mechanical resonator coupled to between one and six damped TLS's that are near-resonant with the resonator,  where the latter is initially prepared in either a single Fock state or superposition of  Fock states. We  find, perhaps surprisingly, that the damping and decoherence dynamics resembles quite closely that for the Ohmic oscillator bath model even with only three near resonant damped TLS's furnishing the mechanical resonator environment. In particular, the Fock state lifetime is observed to scale closely as $1/n$, where $n$ is the initial number of resonator quanta (Fock state number), while the decoherence time of a superposition of  ground and excited Fock states is found to be close to twice the decay time of the excited state, both in accord with the Ohmic model.  A partial understanding of these numerical results can be obtained from a simpler, Born-Markov approximated master equation model for the resonator subsystem  that treats perturbatively the coupling between the resonator and damped TLS's to second order (with the latter traced over as the bath)  and which facilitates analytical calculations for the decay times. However, even more surprising is the observation that completely removing the TLS's  damping does not alter the Ohmic trends,  even though the Born-Markov master equation model is no longer valid. The latter observation is reminiscent of recent numerical investigations to establish subsystem thermalization of closed, many-body quantum systems.\cite{rigol08}

In the next section we present our model system-environment master equation, with a more detailed derivation given in Ref.~[\onlinecite{remus09}].  Sec.~\ref{sec:nointeract} investigates the damping dynamics of  Fock states and decoherence dynamics of superpositions of Fock states for a mechanical resonator coupled to first a single TLS, then three near-resonant TLS's, and finally six near-resonant TLS's, where direct interactions between the TLS's are neglected. An approximate master equation model is presented, yielding analytical decay rate expressions that partially explain the numerically observed trends.  In  Sec.~\ref{sec:TLS-TLS}, we begin with deriving the oscillator-TLS Hamiltonian with pairwise interactions between TLS's taken into account.     The effect of  TLS-TLS interactions on the damping of Fock states and decoherence of Fock state superpositions in a resonator coupled to first three and then six near-resonant, interacting TLS's is investigated.  Finally, we offer some concluding remarks in Sec.~\ref{sec:conclusion}.  

\section{\label{sec:model}Resonator-TLS System Equations}

In this section we present the model for the resonator-TLS system.  For the TLS Hamiltonian we have
\begin{equation}
\hat{H}_{\mathrm{TLS}}=\sum_{\alpha=1}^N \left[ \frac{1}{2}\Delta_0^{(\alpha)}\sigma_z^{(\alpha)}+\frac{1}{2}\Delta_b^{(\alpha)}\sigma_x^{(\alpha)}\right],
\label{TLSHamiltonianeq}
\end{equation}
where $\alpha=1, 2,...,N$ labels the TLS,  $\Delta^{(\alpha)}_0$  is the asymmetry of the $\alpha$th TLS's  potential well and $\Delta_b^{(\alpha)}$ is its tunnel splitting that depends on the well barrier height and width.  Writing out the full oscillator-TLS system Hamiltonian, we have
\begin{equation}
\hat{H}_S=\hbar\omega ({a}^\dagger{a}+1/2) +\sum_{\alpha=1}^N \left[ \frac{1}{2}\Delta_0^{(\alpha)}\sigma_z^{(\alpha)}+\frac{1}{2}\Delta_b^{(\alpha)}\sigma_x^{(\alpha)}+\lambda^{(\alpha)} ({a}+{a}^\dagger)\sigma_z^{(\alpha)}\right],
\label{fullhamiltonianeq}
\end{equation}
where ${a}^\dagger$ and ${a}$ are raising and lowering operators for the resonator mode of interest, satisfying the commutation relation $[{a},{a}^\dagger]=1$.  The strength of the coupling $\lambda^{(\alpha)}$ depends on the location of the TLS defect within the resonator.

In Ref.~[\onlinecite{remus09}] we derive the following master equation describing the dissipative dynamics of the coupled resonator-TLS system: 
\begin{eqnarray}
\dot{\rho}_S (t)&=&-\frac{i}{\hbar}[H_S,\rho_S(t)]-\frac{i\gamma}{2\hbar} [Y,\{ P_Y,\rho_S(t)\}]-\frac{m\omega\gamma}{2\hbar}\coth\left(\frac{\hbar\omega}{2k_B T}\right)[Y,[Y,\rho_S(t)]]\cr
&&-\sum_{\alpha=1}^N\frac{1}{4T_1^{(\alpha)}}\left(\frac{E^{(\alpha)}}{\Delta^{(\alpha)}_b}\right)^2[\sigma^{(\alpha)}_z,[\sigma^{(\alpha)}_z,\rho_S(t)]]
\cr&&-\sum_{\alpha=1}^N\frac{i}{4T_1^{(\alpha)}}\left(\frac{E^{(\alpha)}}{\Delta^{(\alpha)}_b}\right)\tanh\left(\frac{E^{(\alpha)}}{2 k_B T}\right)[\sigma^{(\alpha)}_z,\{\sigma^{(\alpha)}_y,\rho_S(t)\}],
\label{quantumBMMastereq}
\end{eqnarray}
where $\rho_S(t)$ is the resonator-TLS system density matrix, $Y=Y_{zp}(a+a^{\dagger})$ gives the mechanical resonator mode displacement, with $Y_{zp}$ the zero-point displacement uncertainty,  $P_Y$ is the resonator mode momentum, and $E^{(\alpha)}=\sqrt{(\Delta_0^{(\alpha)})^2+(\Delta_b^{(\alpha)})^2}$ is the $\alpha$th TLS energy level separation. The parameter $\gamma$ gives the energy damping rate of the oscillator in the absence of the TLS's, while the parameter $T_1^{(\alpha)}$ gives the $\alpha$th TLS relaxation time from its excited energy eigenstate in the absence of the oscillator.  We shall use dimensionless time units, $t\rightarrow\omega t$, with  $T_1$ and $\gamma$  expressed as $\omega T_1$ and  $\gamma/\omega$, respectively,  and $\lambda$, $\Delta_i$, and temperature $T$  expressed as $\lambda/\hbar\omega$, $\Delta_i/\hbar\omega$, and $k_BT/\hbar\omega$, respectively.

\section{\label{sec:nointeract}Damping and Decoherence Due to Non-Interacting TLS's}     

\subsection{\label{sec:1fockdamping}Single TLS}
\begin{figure}[htbp]
	\centering
		\includegraphics[height=2.2in]{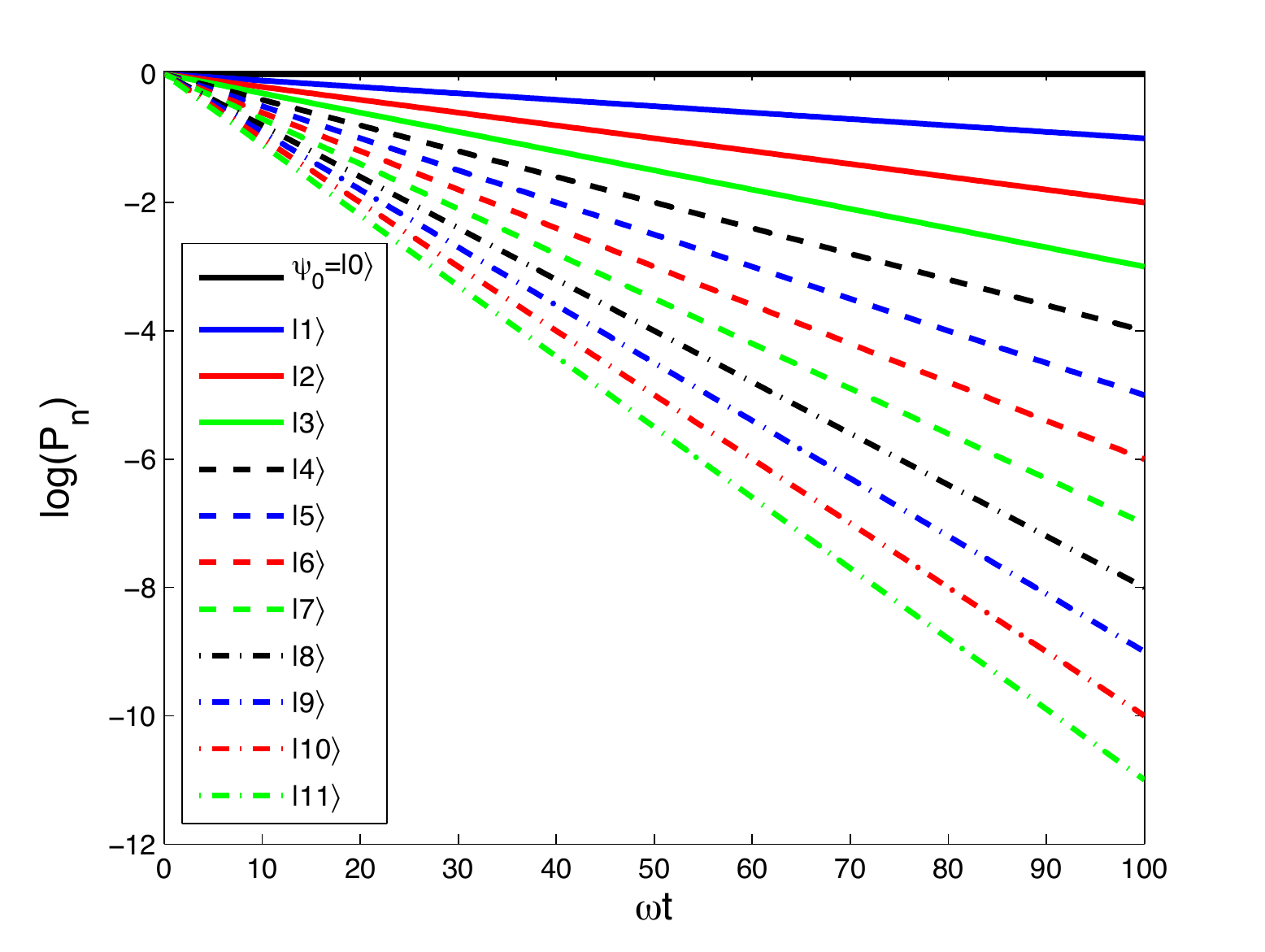}
	\caption{(Color online) Log of number state probability $P_n$ vs $\omega t$ for a range initial Fock states for the resonator coupled to an Ohmic oscillator bath only, where $\gamma=0.01$ and $T=0.09$.}
	\label{P_n_no_spin}
\end{figure}

In this section we investigate the damping of Fock states and the decoherence of Fock state superpositions in a mechanical resonator interacting with a single damped TLS. As a partial check of our numerical methods,  we begin by evaluating the number state probability $P_n=\langle n|\rho|n\rangle$ as a function of time for a resonator mode coupled to an Ohmic oscillator bath only, where the analytical solution is known.  Fig.~\ref{P_n_no_spin} shows the log of the number state probability for initial Fock states $|n\rangle$, with $n=0$ to $n=11$, in the absence of the damped TLS. The slope of each successive curve decreases by an increment of $1/T_{11}$, where $T_{11}$ is the lifetime of the first excited state; as expected, the number state  lifetime decays as $1/n$.\cite{lu89} 

\begin{figure}[htbp]
	\centering
		\includegraphics[height=2.2in]{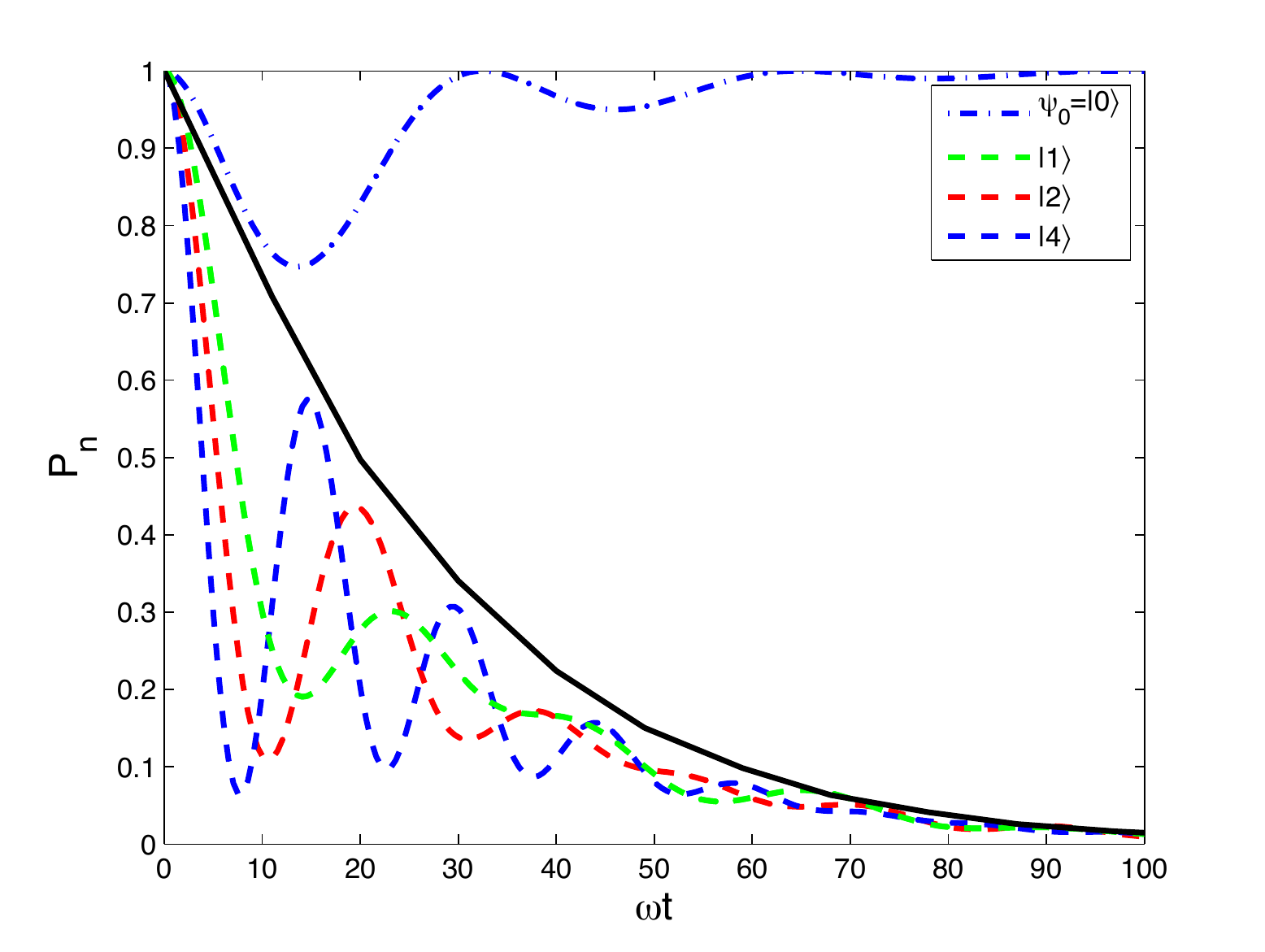}
		\includegraphics[height=2.2in]{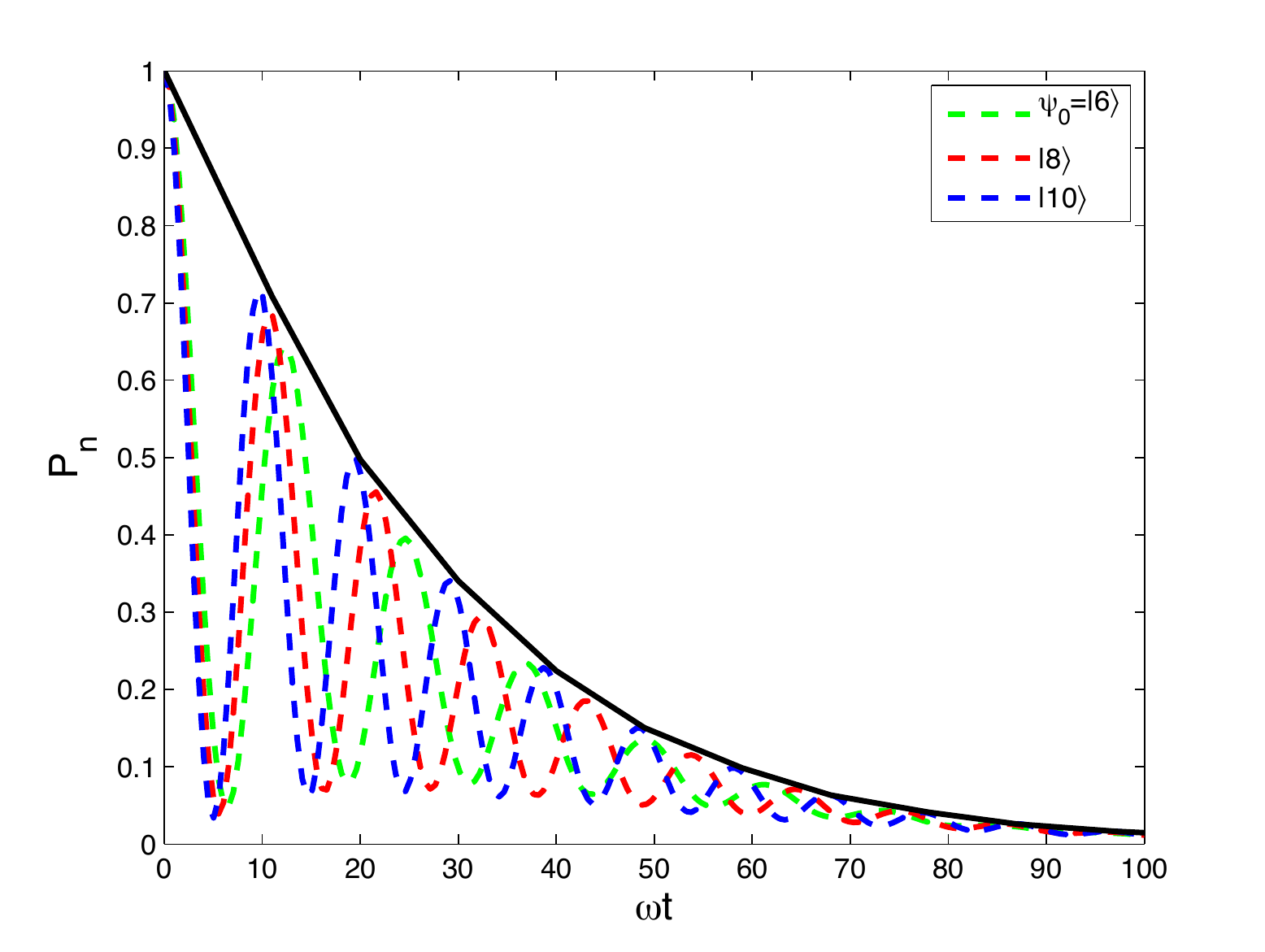}
	\caption{(Color online) Number state probability $P_n$ vs $\omega t$ for various initial Fock states for the resonator coupled to a damped TLS only. The black curve, indicating the peaks of the curves in the right-hand plot, is the same in both plots.  For both plots $\Delta_0=0$, $\Delta_b=1$, $\lambda=0.1$, $T=0.09$, and $T_1=10$.}
	\label{small_and_large_n}
\end{figure}

Fig.~\ref{small_and_large_n} shows the number state probability for the resonator coupled  to a single, on-resonance damped TLS.  In this case we see that the $P_n$ curves oscillate, as energy is transferred from the resonator to the TLS and back. As a partial check of the numerics, the time of the first minimum of each curve corresponds closely  to the Jaynes-Cummings model prediction for the transfer time of a quantum of vibrational energy to a symmetric, on resonance TLS:  $\omega t=\pi E/(2\lambda\sqrt{n})$, where $\hbar \omega=E=\Delta_b$.     The left-hand plot shows the number state probability for four low-$n$ states, while the right-hand plot displays curves for higher energy states.  The three high-$n$ states all appear to decay at the same rate, as indicated by the black curve, which simply follows the maxima of the undulating curves.  The same black curve is shown in the left-hand plot, and in this case the $P_n$ plots clearly fall short of this ``maximum" curve.  Further, the extent to which the curves fall short depends on their $n$ value, with the $n=1$ curve having the lowest amplitude.  Thus, for low-energy Fock states the Fock state probability appears to exhibit $n$-dependent damping, while for higher-$n$ states the damping does not depend on the number state.  This $n$-dependence may be an indication of TLS saturation.  For low-$n$ Fock states the unsaturated TLS pulls energy from the resonator, where it is then dissipated to the TLS bath, causing the number state probability to decay more quickly.  For higher-$n$ states, however, the TLS is saturated and thus contributes uniformly to damping, independent of the initial $n$.       

\begin{figure}[htbp]
	\centering
	\subfigure[]{
		\includegraphics[height=2.1in]{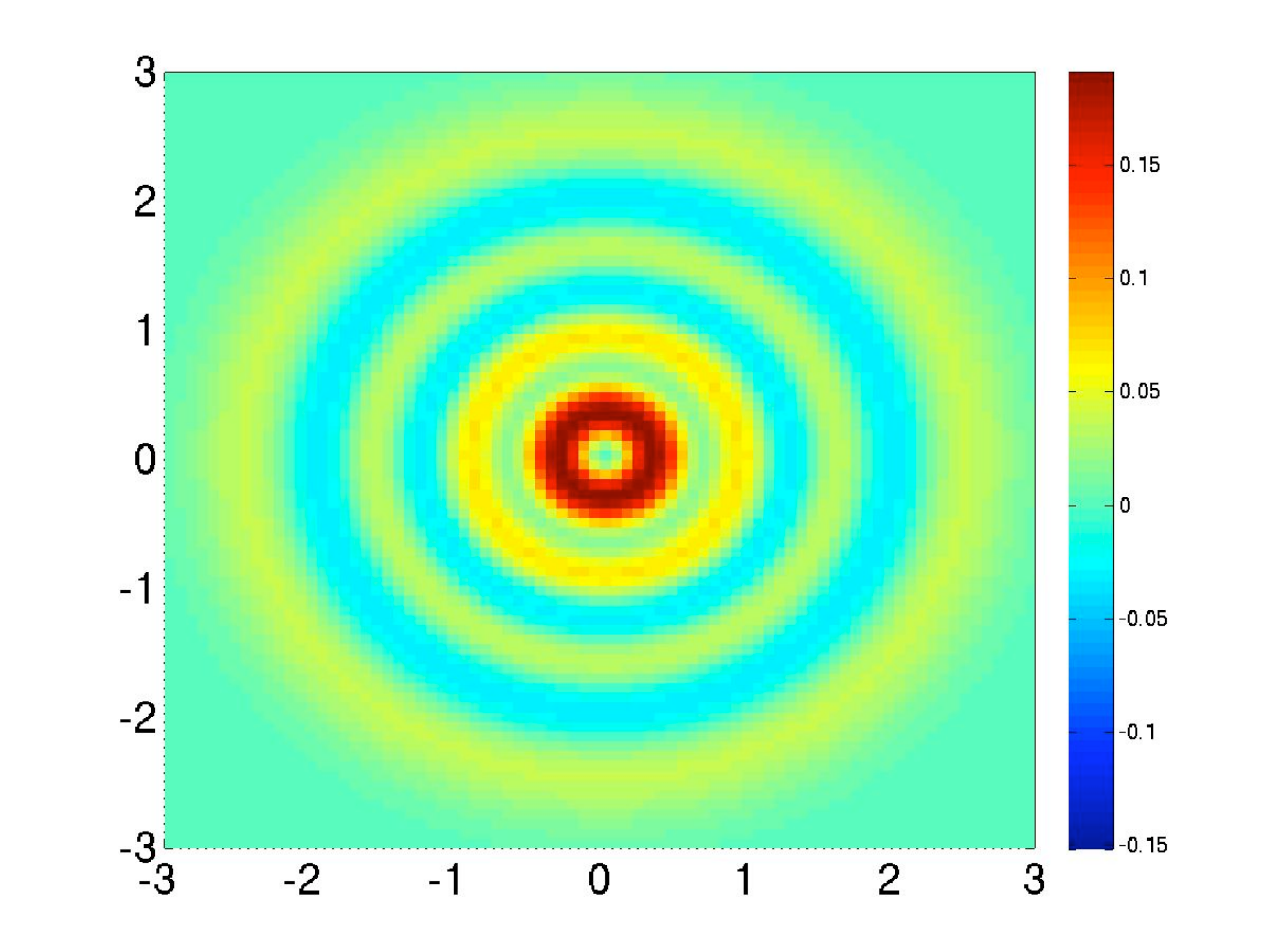}
	\label{snapshot_mixed}
	}
	\subfigure[]{
		\includegraphics[height=2.1in]{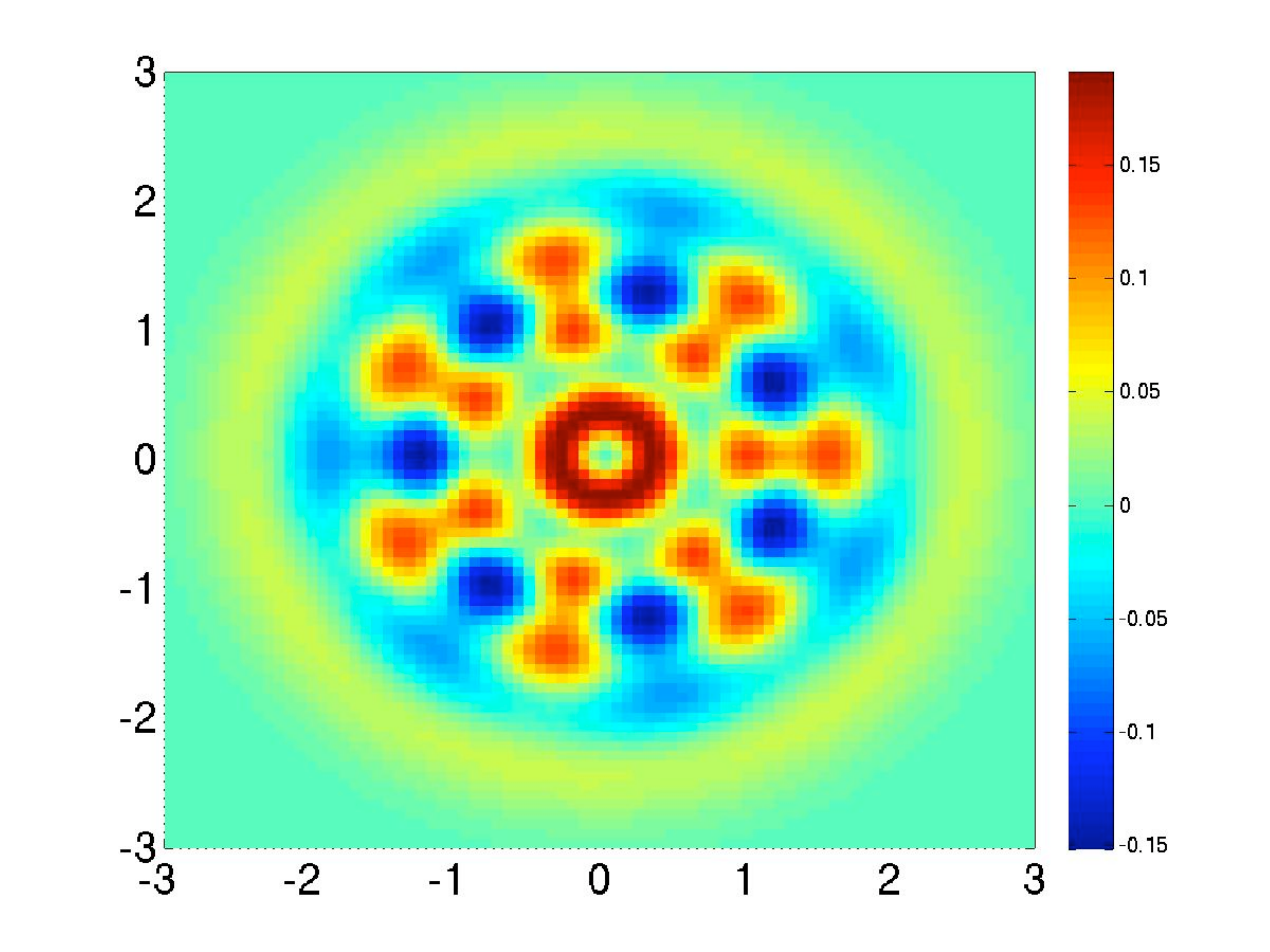}
	\label{snapshot_super}
	}
	\caption{(Color online) Wigner function for a mixture of number states $|0\rangle$ and $|7\rangle$ (a) and for a superposition of the same states (b).} 
	\label{initial_states}
\end{figure}

\begin{figure}[htbp]
	\centering
		\includegraphics[height=1.35in,width=1.45in]{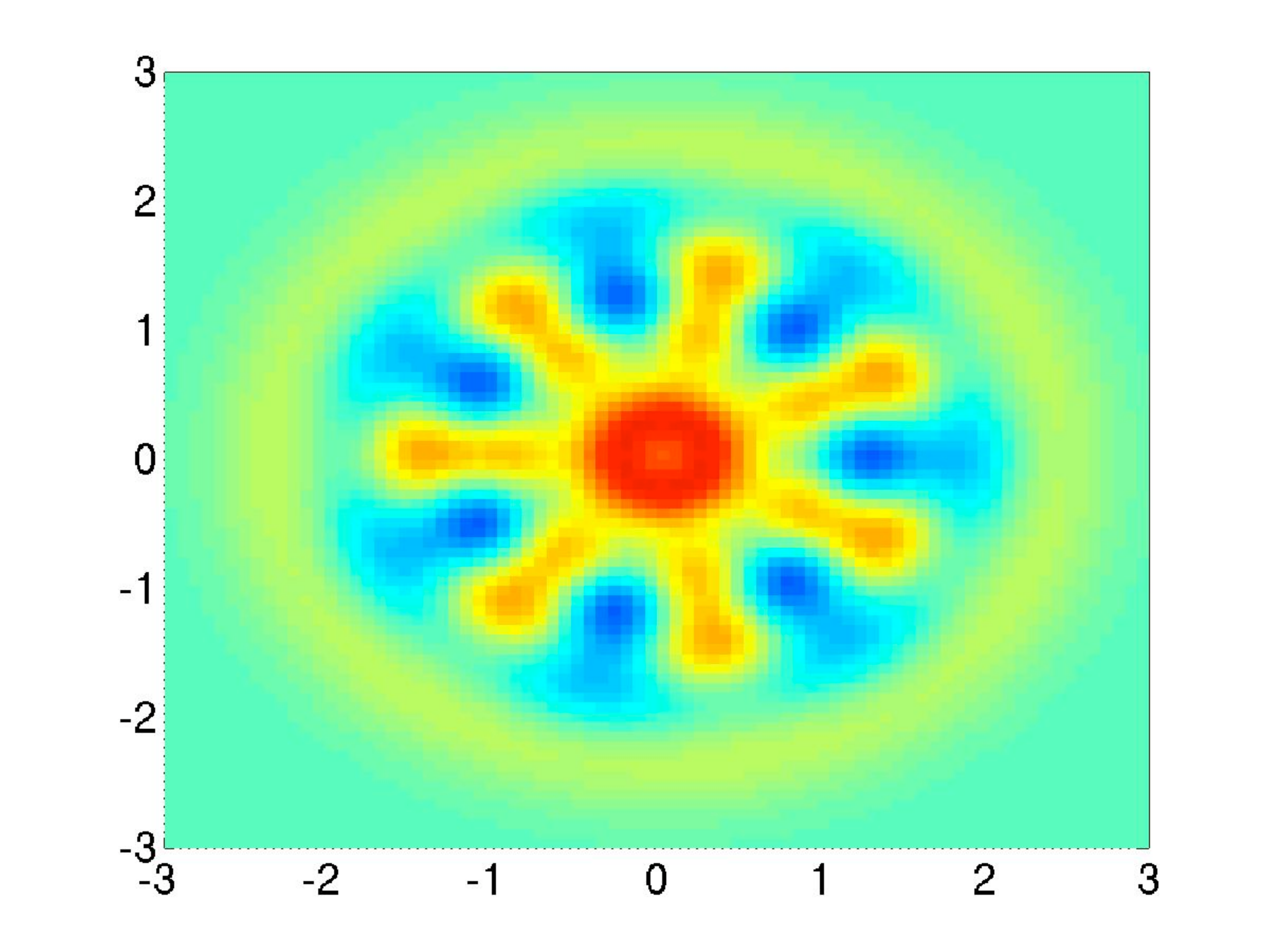}
		\includegraphics[height=1.35in,width=1.45in]{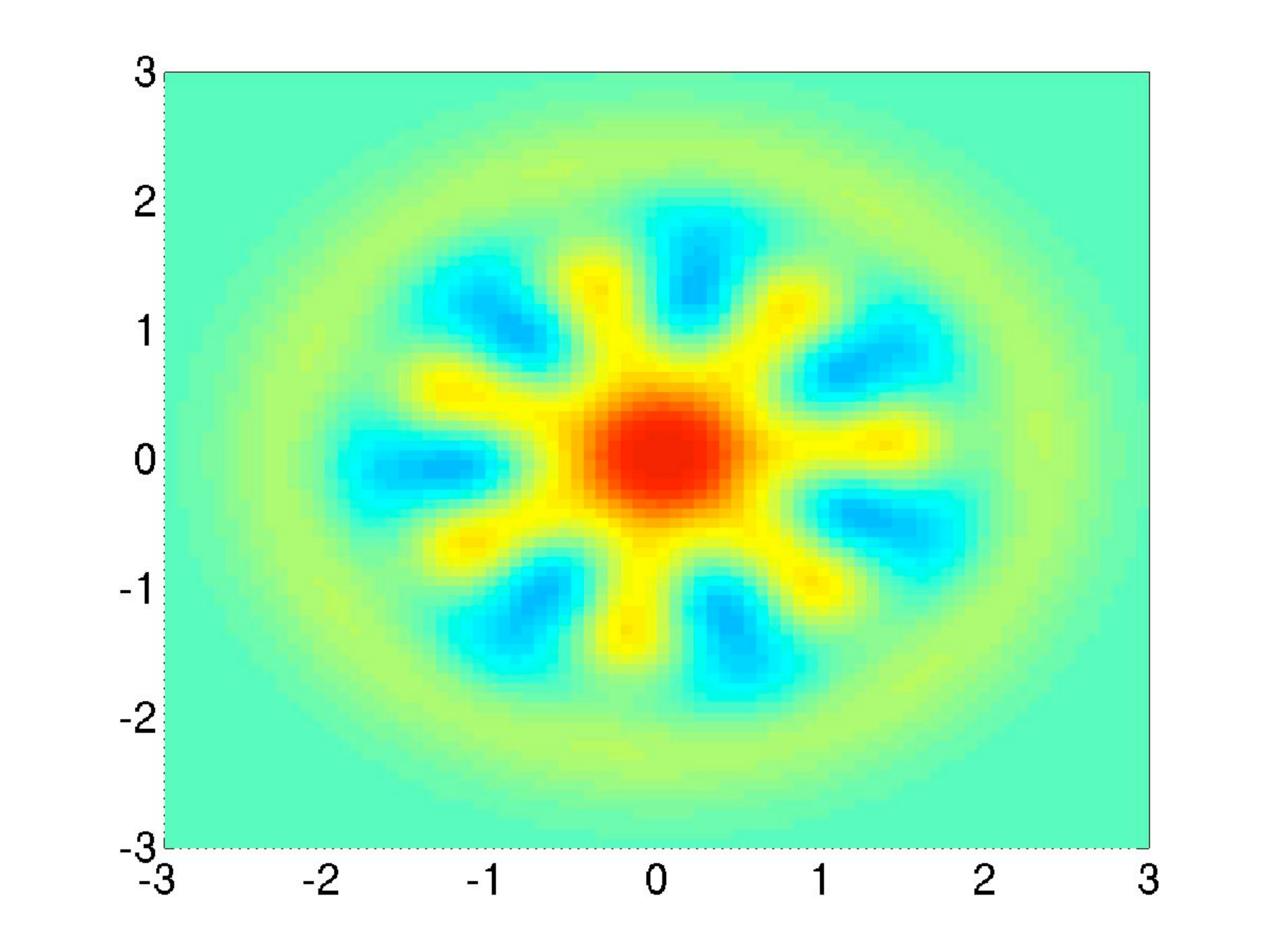}
		\includegraphics[height=1.35in,width=1.45in]{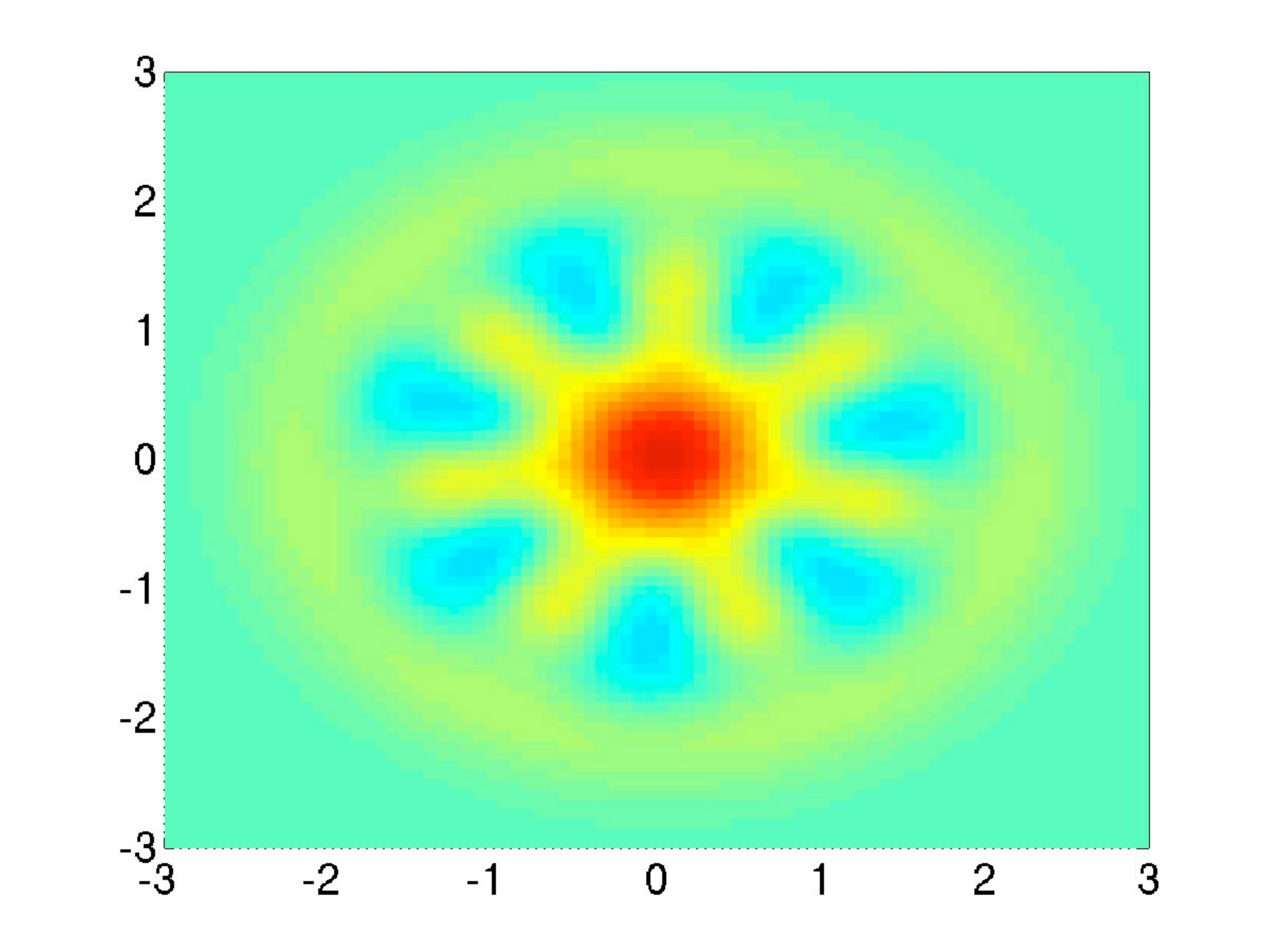}
		\includegraphics[height=1.35in,width=1.45in]{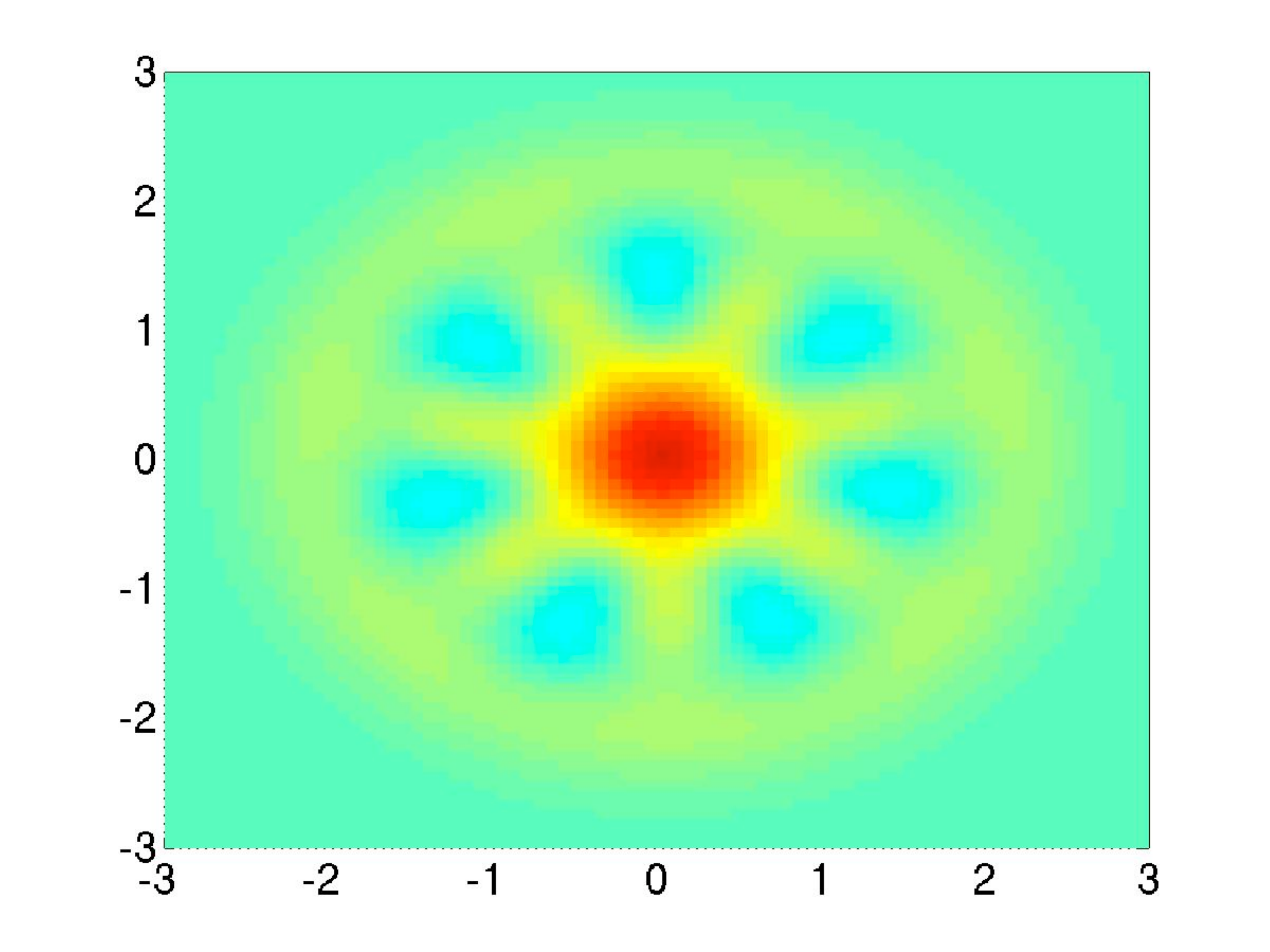}
	\caption{(Color online) Evolving Wigner function for the resonator  initially in the superposition state shown in Fig.~\ref{snapshot_super} coupled to an Ohmic oscillator bath with $\gamma=0.01$ and $T=0.09$.}
	\label{dephasing:osc}
\end{figure}

\begin{figure}[htbp]
	\centering
		\includegraphics[height=1.35in,width=1.45in]{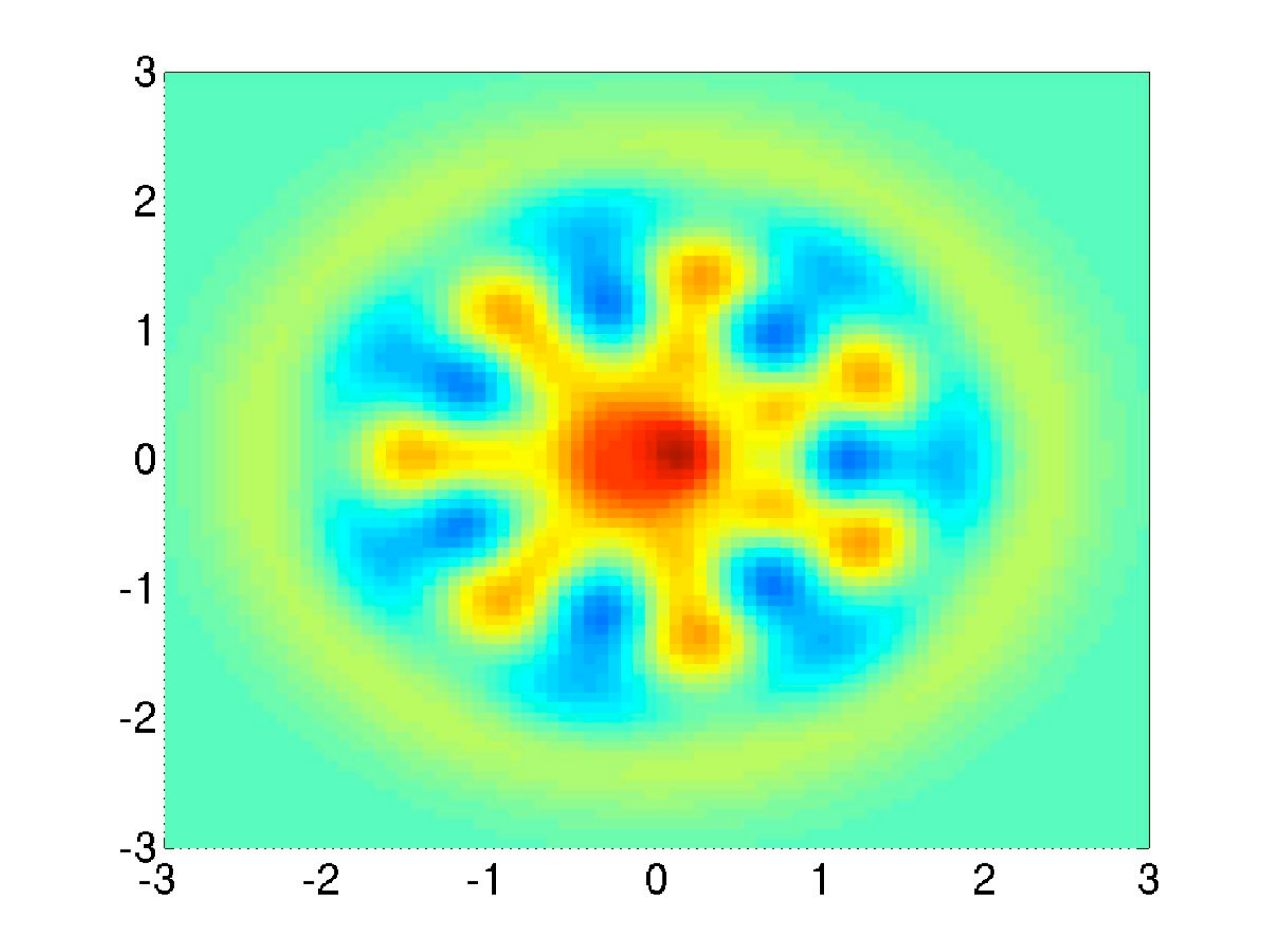}
		\includegraphics[height=1.35in,width=1.45in]{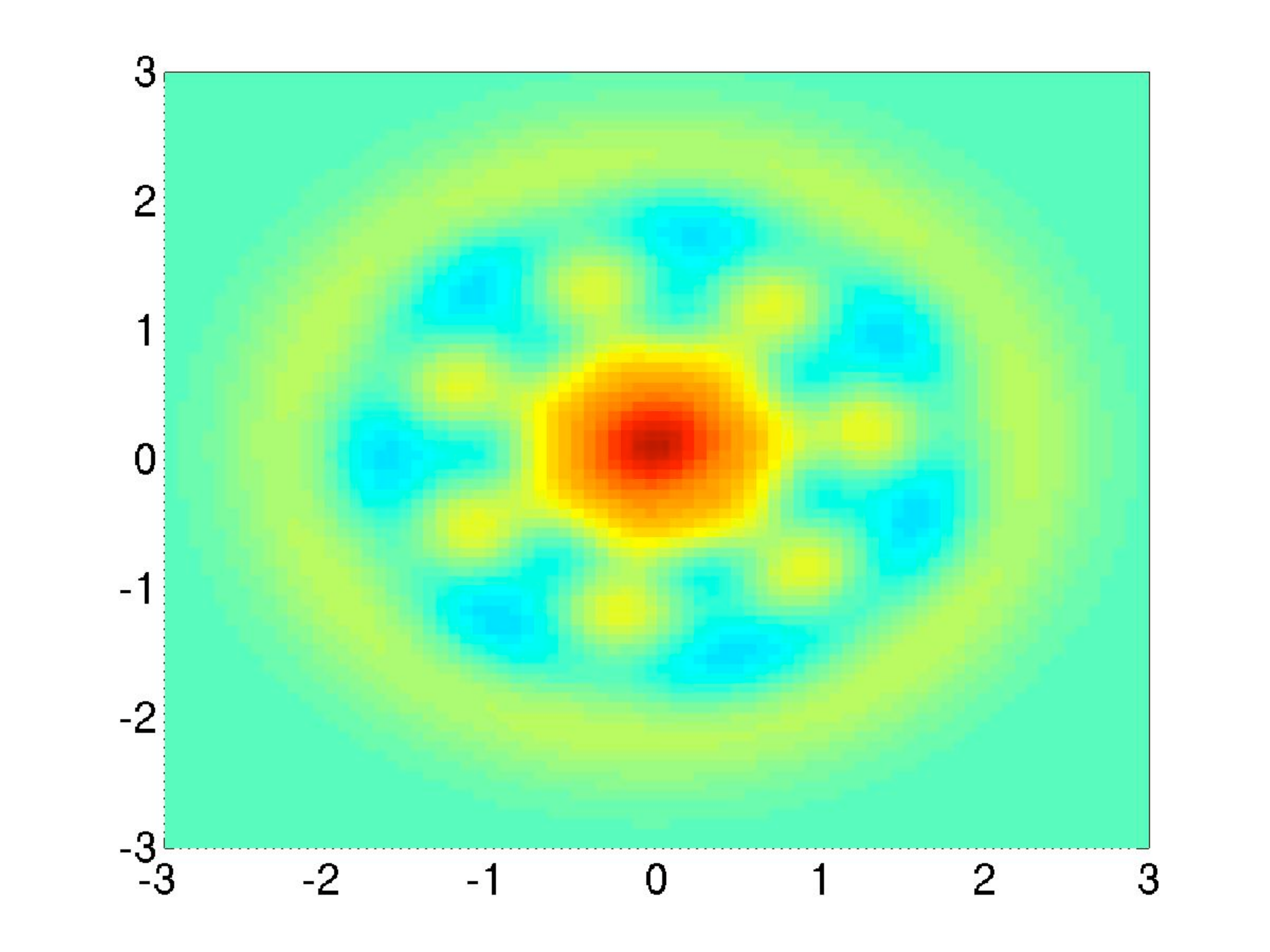}
		\includegraphics[height=1.35in,width=1.45in]{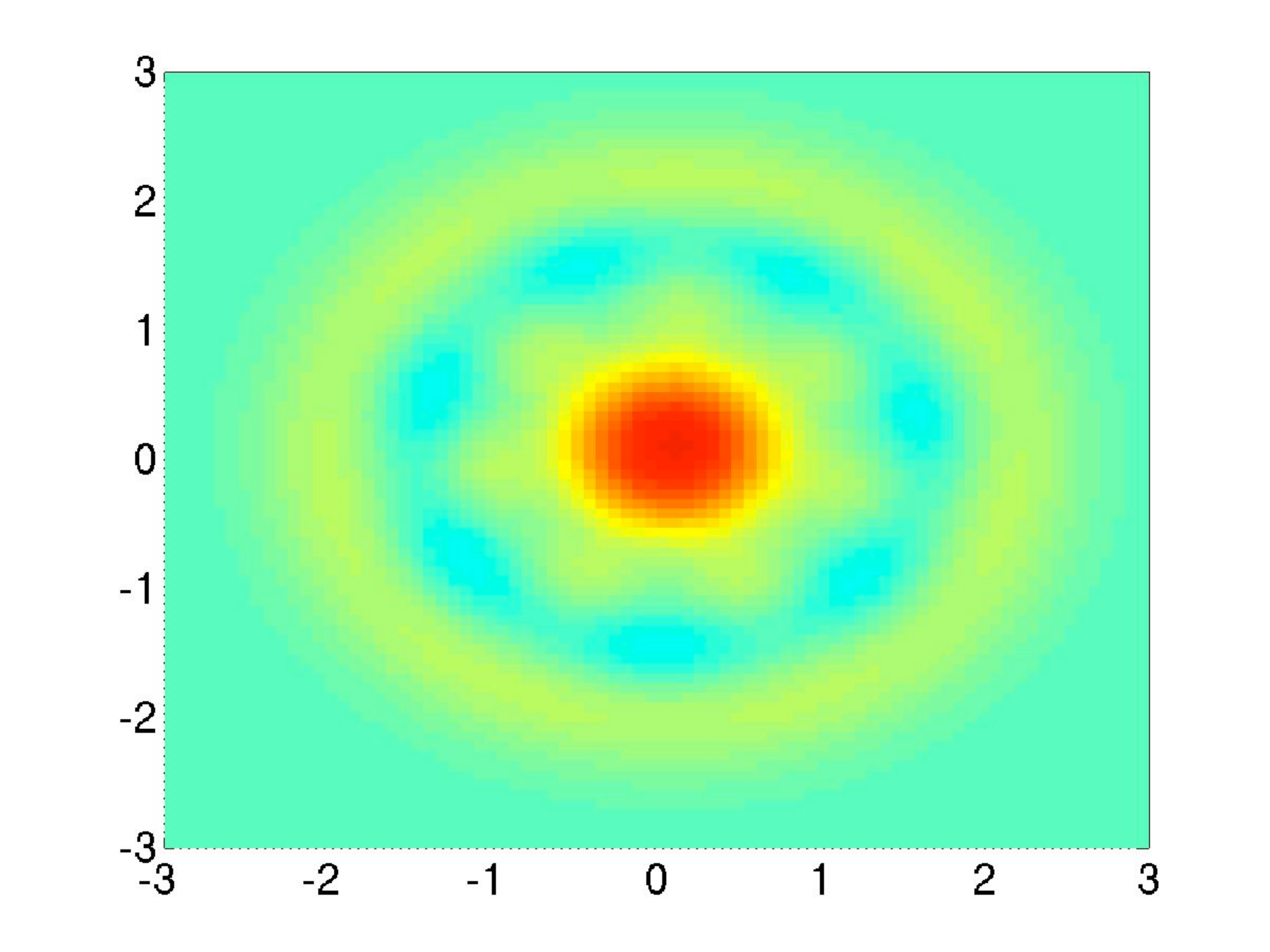}
		\includegraphics[height=1.35in,width=1.45in]{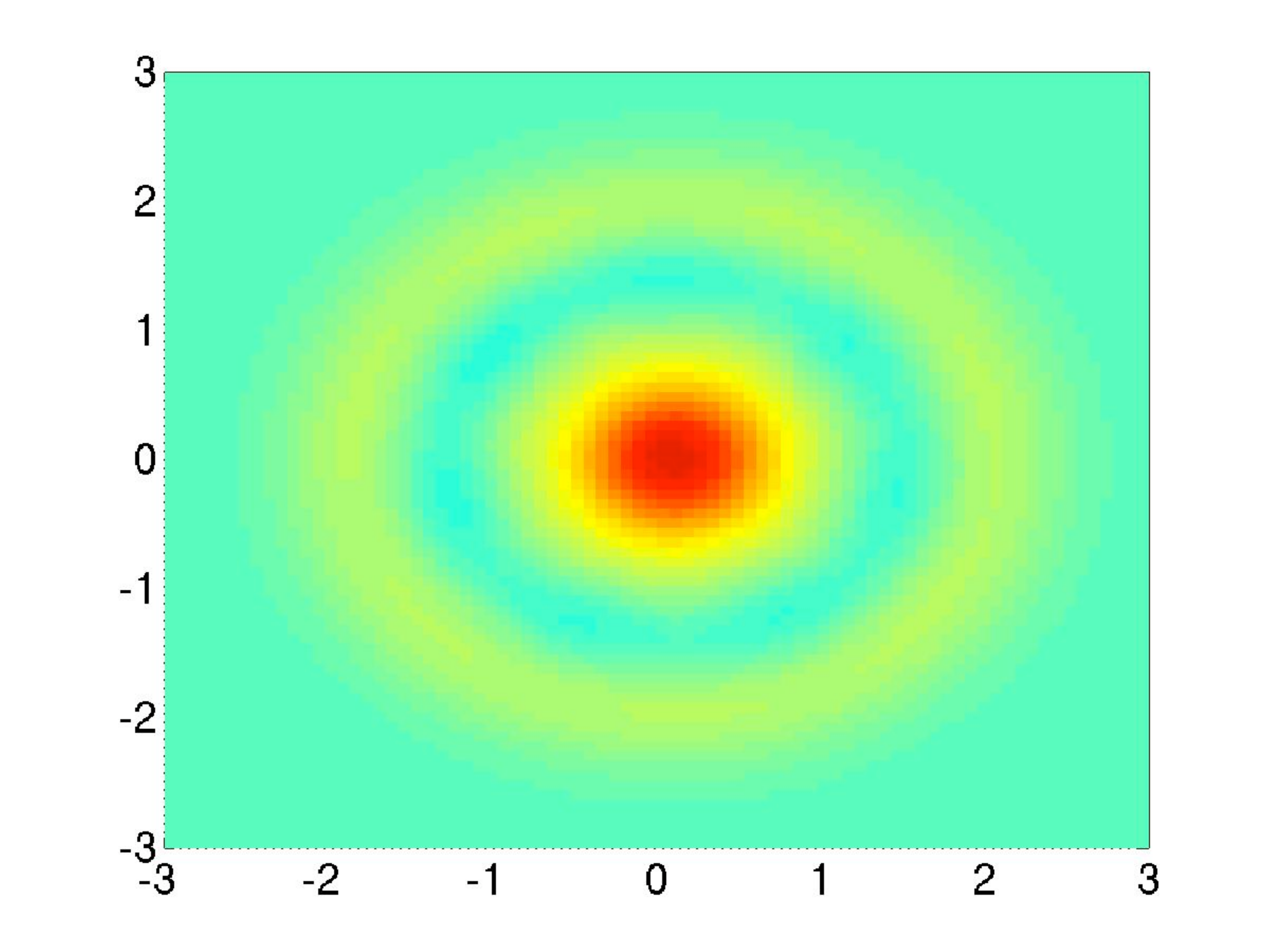}
	\caption{(Color online) Evolving Wigner function for the resonator  initially in the superposition state shown in Fig.~\ref{snapshot_super} coupled to a damped TLS with $\lambda=0.1,\Delta_0=0,\Delta_b=1,T=0.09,\mathrm{and}T_1=10.$}
	\label{dephasing:spin}
\end{figure}

Next, we investigate the effect of a damped TLS on number state superpositions. A useful `visual' representation of the state is its Wigner function;  Fig.~\ref{initial_states} shows two initial oscillator states: an equal mixture of the ground and $n=7$ state [Fig.~\ref{snapshot_mixed}], and a superposition of the same two Fock states [Fig.~\ref{snapshot_super}]. In both cases the Wigner function has positive and negative values, because both the Fock state mixture and the superposition are non-classical states.  However, the spoke-like interference fringes in the superposition plot indicate the presence of non-zero off-diagonal terms of the density matrix, as opposed to the concentric undulations in the mixture plot.  
Fig.~\ref{dephasing:osc} shows four equally-spaced snapshots of the Wigner function for a resonator initially in the superposition state shown in Fig.~\ref{snapshot_super}.  The resonator is coupled to an Ohmic bath that causes the state to decay and the amplitude of the Wigner function to decrease.  However, both the ring and spoke-like structures of the initial state are still visible in the final snapshot.  Fig.~\ref{dephasing:spin} shows a similar set of snapshots, this time for a resonator coupled to an on-resonance, damped TLS only.  
In contrast to the superposition state decay in the Ohmic bath case, we see that the spoke-like structure disappears first, leaving concentric rings similar to those seen in Fig.~\ref{snapshot_mixed}.  The dephasing time $T_{\phi}$ is usually defined in terms of the decay times of the on- and off-diagonal terms of the resonator's density matrix as follows:
\begin{equation}
\frac{1}{T_{0n}}=\frac{1}{2T_{nn}}+\frac{1}{T_{\phi}},
\label{Tphieq}
\end{equation}
where $T_{0n}$ is the lifetime of the off-diagonal density matrix element $\rho_{0n}$, and $T_{nn}$ is the lifetime of the diagonal matrix element $\rho_{nn}$.  The disappearance of the spokes prior to the rings suggests a finite $T_{\phi}$, in contrast to an oscillator bath, where $T_{0n}=2T_{nn}$.

\subsection{\label{sec:3noint}Three TLS's}

We now increase the number of damped TLS's to three.   The TLS energies  $\Delta^{(\alpha)}_{0}$ and $\Delta^{(\alpha)}_{b}$, $\alpha=1,2,3$,  are chosen randomly according to the Standard Tunneling Model (STM) distribution.\cite{esquinazi98,remus09}   As our condition for near resonance, the corresponding TLS energies $E^{(\alpha)}$  are restricted to the  range  $0.75\hbar\omega\leq E^{(\alpha)}\leq1.25\hbar\omega$, where recall $E^{(\alpha)}=\sqrt{(\Delta_0^{(\alpha)})^2+(\Delta_b^{(\alpha)})^2}$.  We also choose random values for the $T^{(\alpha)}_1$ relaxation times of each individual TLS by first selecting a reference $T_1$ value  and then assigning to each TLS a randomly-generated $T^{(\alpha)}_{1}$ within $\pm50\%$ of the reference value. Furthermore, each TLS is assigned a random $\lambda^{(\alpha)}$ coupling that is within $\pm50\%$ of a reference value $\lambda=0.1/6$,  scaled down from the single TLS coupling considered in the previous section ($\lambda=0.1$) so as to avoid significant TLS-induced renormalizations of the resonator's harmonic potential resulting from having more coupled TLS's. We choose a temperature $T=0.09$ for all plots. 

\begin{figure}[htbp]
	\centering
		\includegraphics[height=2.2in]{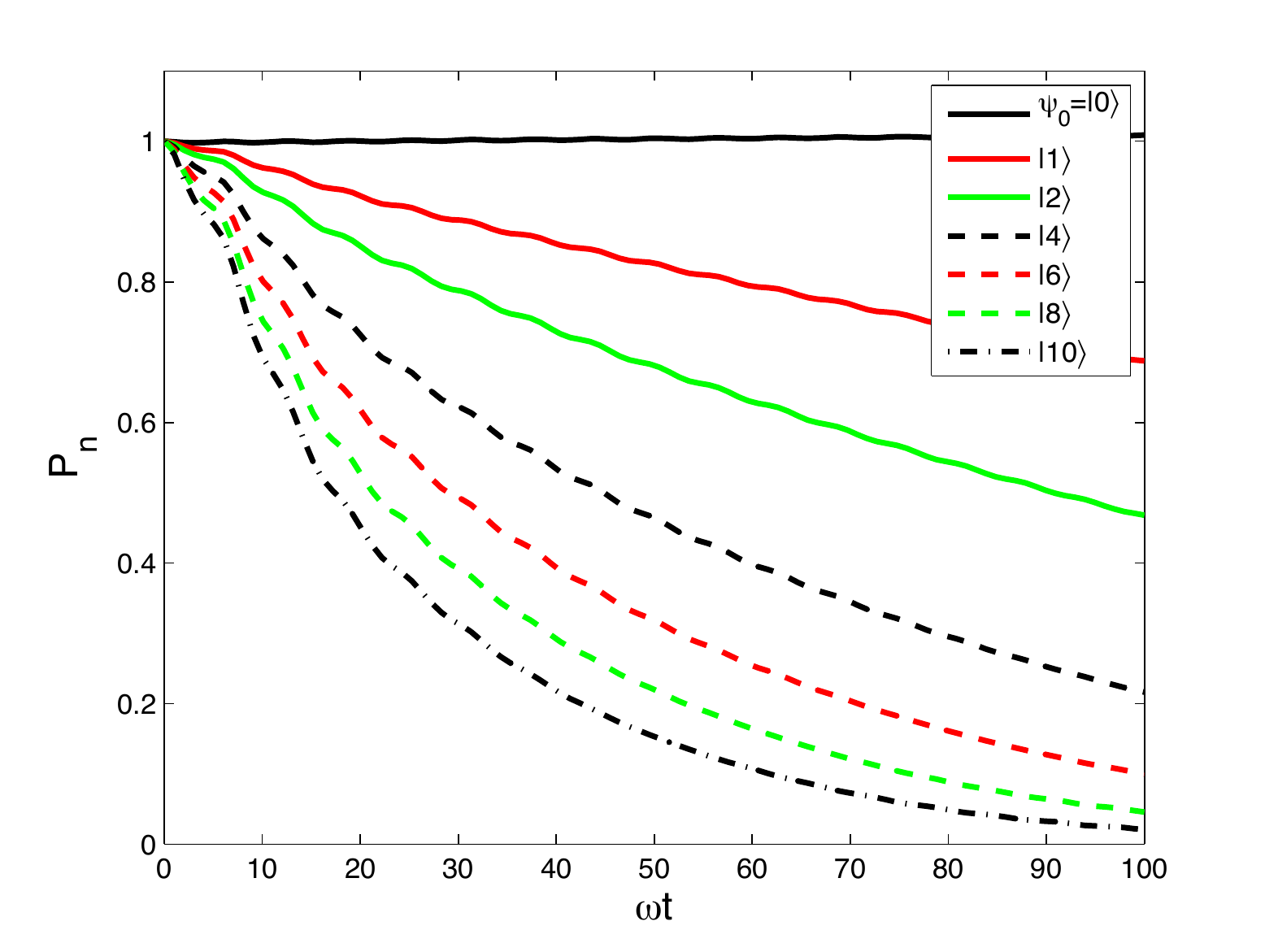}
		\includegraphics[height=2.2in]{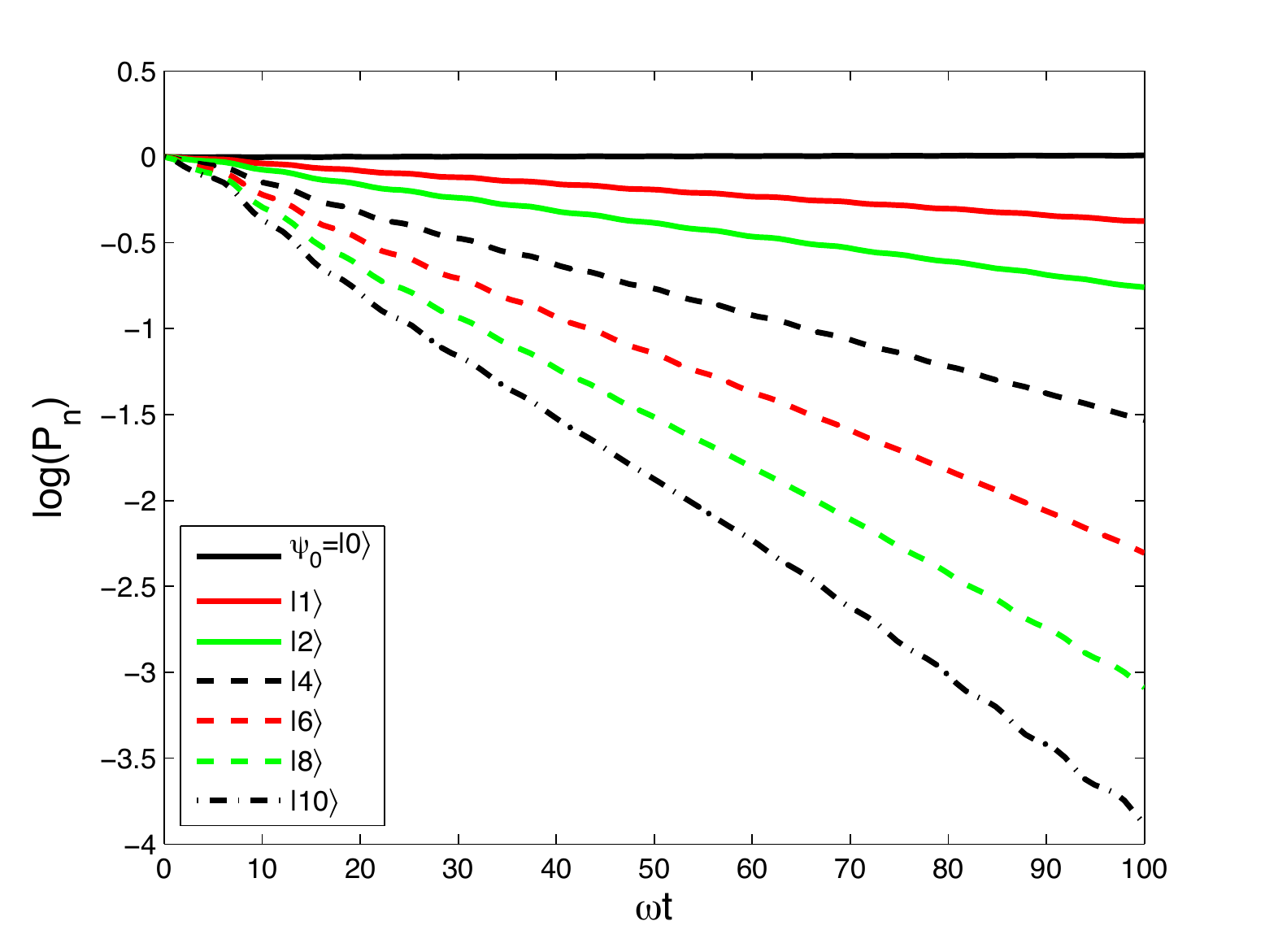}
	\caption{(Color online) Left: $P_n$ vs $\omega t$ for a resonator coupled to three non-interacting TLS's.  Right: Log of $P_n$ vs $\omega t$.  For all curves $T_1=10$.}
	\label{P_n3spins_no_int}
\end{figure}

To investigate Fock state decay, we choose an initial Fock state $|\psi_0\rangle=|n\rangle$ and then determine the corresponding number state probability $P_n$ as a function of time.  Fig.~\ref{P_n3spins_no_int} shows the decay of $P_n$ for a range of initial Fock states.  
In contrast to the single-TLS case, the number state probabilities do not show large oscillations but instead decay relatively smoothly. Furthermore,  the nearly-linear curves in the log plot indicate that  $P_n$ decays exponentially and the decay rates can be extracted from a linear fit.
\begin{figure}[htbp]
	\centering
		\includegraphics[height=3in]{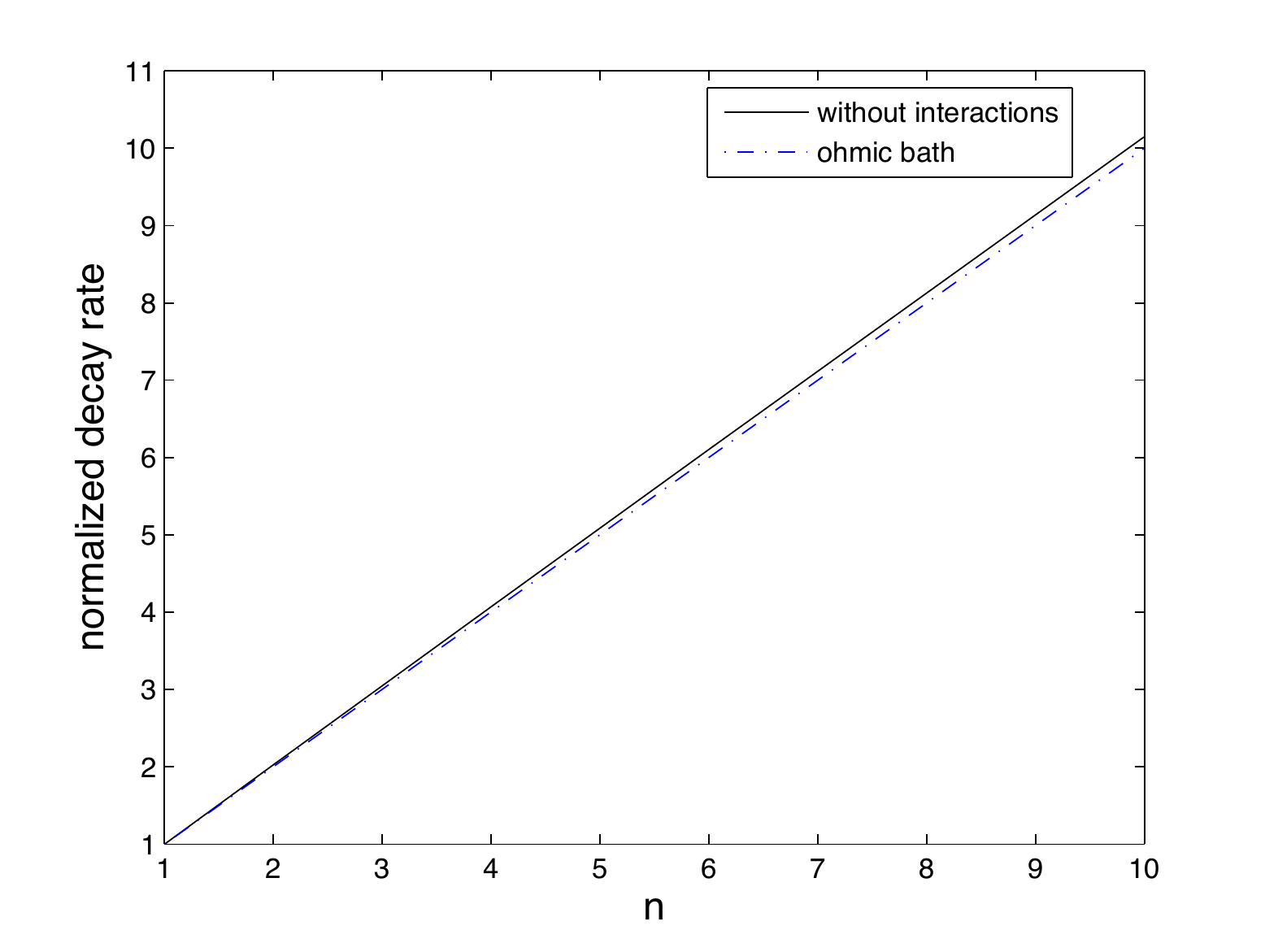}
	\caption{(Color online) Normalized decay rate vs $n$ for single Fock states. The resonator is coupled to three non-interacting TLS's (solid) and to an Ohmic bath without any TLS's present (dot-dash).  For both curves $T_1=10$.}
	\label{n_depend_no_int}
\end{figure}
We noted in the previous section that for a  resonator damped by an Ohmic bath, the decay time for the $n^{th}$ state goes as $T_{nn}=T_{11}/n$, where $T_{11}$ is the decay time for the first excited state: the decay rate scales as $n$.  Fig.~\ref{n_depend_no_int} shows the normalized decay rate $T_{11}/T_{nn}$ for  $P_n$ as a function of $n$ .  For a resonator coupled solely to an Ohmic bath, the curve has a slope equal to one.  For a resonator coupled to three TLS's, the slope is very close to one; Fock states decay similarly to a resonator that is Ohmically coupled to a bath of free oscillators.

\begin{figure}[htbp]
	\centering
		\includegraphics[height=3in]{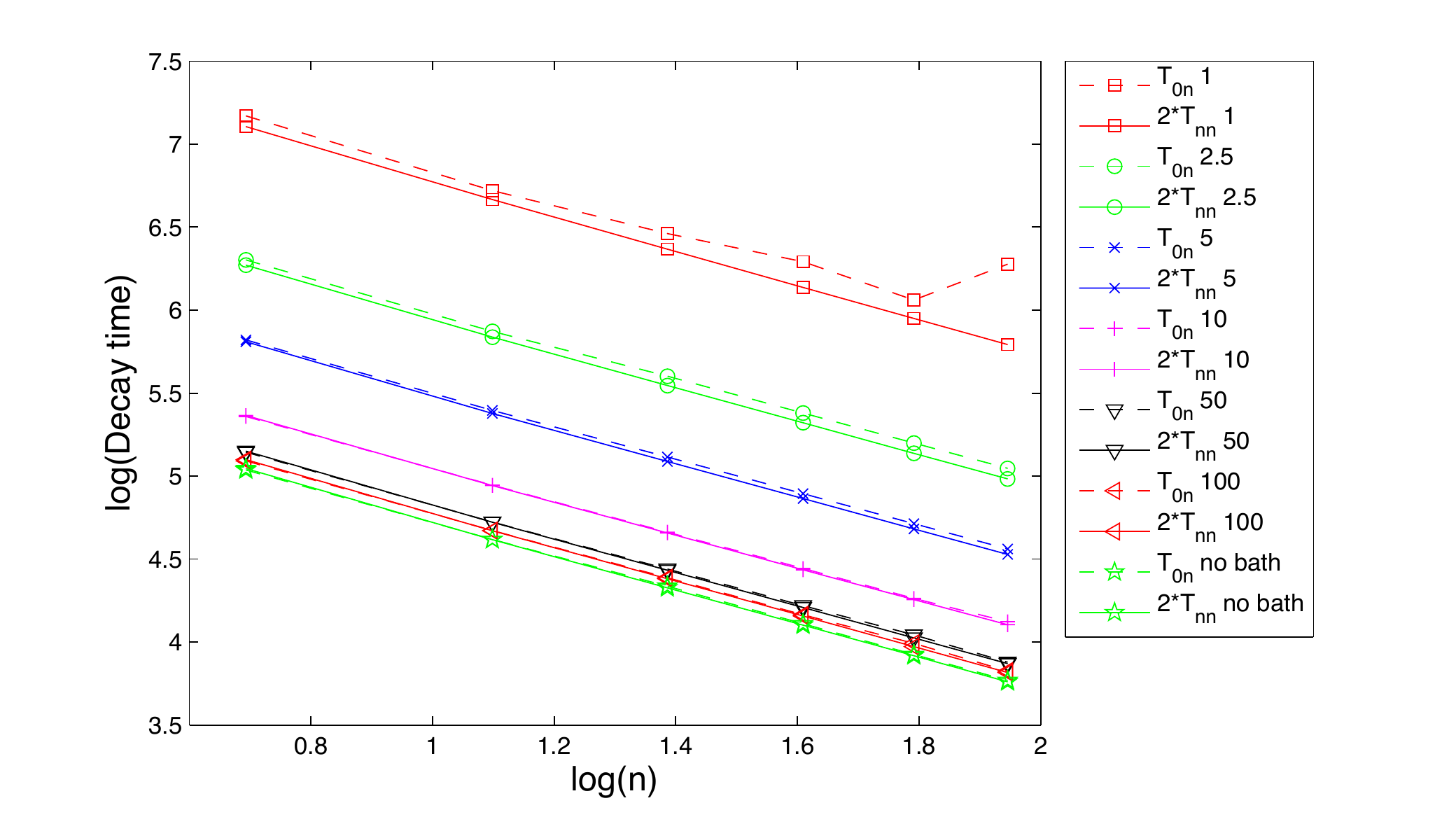}
	\caption{(Color online) $T_{0n}$ and $2T_{nn}$ for the initial superposition state $|\psi_0\rangle=1/\sqrt{2}(|0\rangle+|n\rangle)$ with a range of $T_1$ values shown in the legend.  For all curves the resonator is coupled to three non-interacting TLS's.  For the lowest curve (solid green with star markers) the TLS's are not damped, and thus a $T_1$ time is not given.}
	\label{3spins_no_int}
\end{figure}

We now investigate the decay of a superposition of the ground state and the $n$th excited state, $|\psi\rangle=1/\sqrt{2}(|0\rangle+|n\rangle)$, with each TLS  initially in a thermal state.  We consider the $\rho_{nn}$ and $\rho_{0n}$ elements of the density matrix as a function of time (plots not shown).  The curves  decay approximately exponentially, and the log plots decay linearly.  We thus apply a linear fit to the natural log of the curves to find the diagonal and off-diagonal decay times, $T_{nn}$ and $T_{0n}$, respectively.  Fig.~\ref{3spins_no_int} shows the log of the decay times as a function of $\log(n)$ for a range of $T_1$ values.  We plot $2T_{nn}$ to allow for a comparison to the relation $T_{0n}=2T_{nn}$ for an Ohmic bath.  The curves in Fig.~\ref{3spins_no_int} all decay uniformly and with a slope $\approx-1$.   The $2T_{nn}$ and $T_{0n}$ curves are very similar: dephasing is negligible compared to decay.  

The curves in Fig.~\ref{3spins_no_int} show a surprising dependence on $T_1$.  As a reminder, $T^{(\alpha)}_{1}$ is  the decay time of the $\alpha\mathrm{th}$ TLS  from its excited to ground state.  Because $T_1$ determines the strength of the coupling between a TLS and its bath, with smaller $T_1$ corresponding to stronger coupling, we would expect $T_{nn}$ and $T_{0n}$ to decrease as $T_1$ decreases; stronger coupling would result in shorter resonator Fock state decay times.  However, Fig.~\ref{3spins_no_int} shows that the opposite is true.  The curve with $T_1=1$ shows longer decay times than the curve with $T_1=100$.  The lowest (solid green) curve is for a resonator coupled to three {\emph {undamped}} TLS's, and thus a $T_1$ for this curve is not given.  This curve shows the shortest decay times, and appears to be the large-$T_1$ limit of the curves for the damped TLS's.   

\begin{figure}[htbp]
	\centering
		\includegraphics[height=3in]{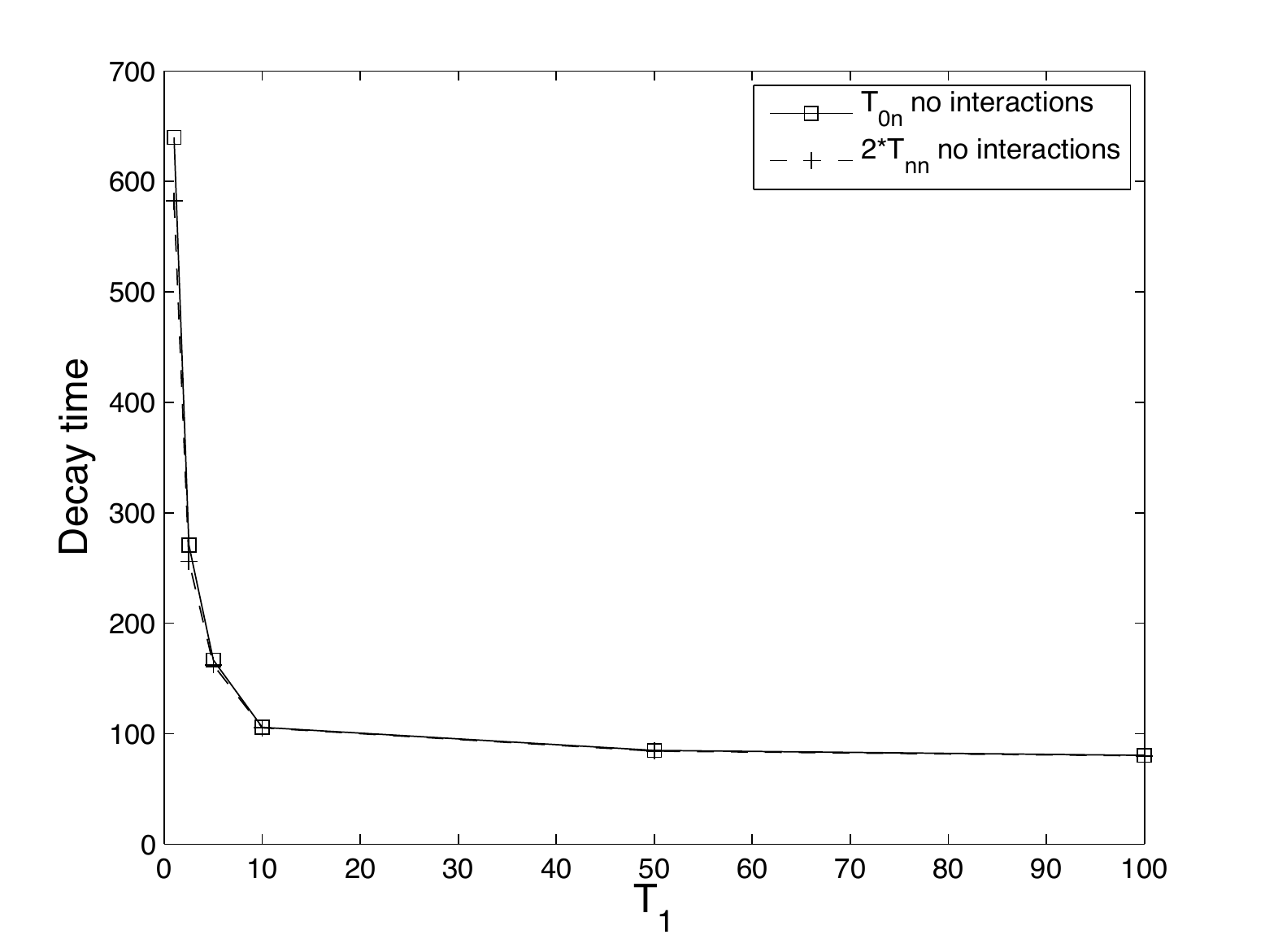}
	\caption{$2T_{nn}$ and $T_{0n}$ vs $T_1$ for the $n=4$ superposition state.  The resonator is coupled to three non-interacting TLS's.}
	\label{T_13spins_no_int}
\end{figure}

\begin{figure}[htbp]
	\centering
		\includegraphics[height=3in]{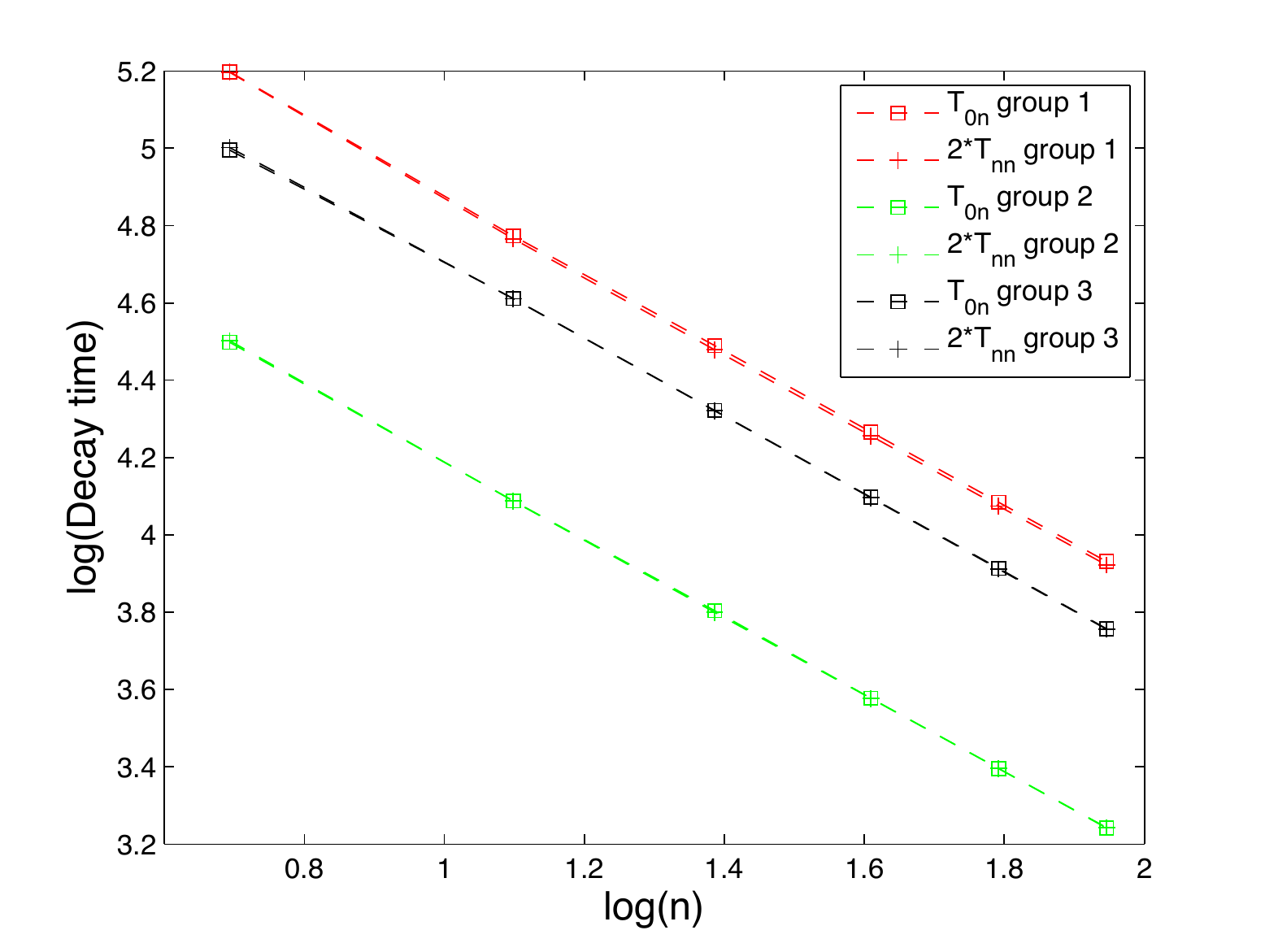}
	\caption{(Color online) $2T_{nn}$ and $T_{0n}$ vs $n$ for three different groups of the TLS parameters.  For all curves $T_1=10$.}
	\label{3groups_3spins_no_int}
\end{figure}

To further investigate the $T_1$ dependence of the decay times, in Fig.~\ref{T_13spins_no_int} we plot $T_{0n}$ and $2T_{nn}$ as a function of $T_1$ for the $n=4$ superposition state.  The plot shows a strong dependence on $T_1$, particularly for $T_1<10$, and suggests that reducing the TLS-bath coupling causes superposition states to decay more quickly.  This surprising dependence on $T_1$ will be discussed in further detail below in Sec.~\ref{sec:approx}.
Finally, in Fig.~\ref{3groups_3spins_no_int} we plot  $2T_{nn}$ and $T_{0n}$ vs $n$ for three different realizations of the randomized TLS parameters.  While the curves indicate the same qualitative linear dependence $T_{nn}= T_{11}/n$, there is some scatter in the  $T_{11}$ values, as indicated by the different intercepts. This is to be expected given that we have only a small statistical sample of three randomly selected TLS's coupled to the resonator.

\subsection{\label{sec:6noint}Six TLS's}

We now consider a resonator coupled to six non-interacting TLS's.  We assign random values to the TLS energies $\Delta^{(\alpha)}_{0}$ and $\Delta^{(\alpha)}_{b}$ according to the STM distribution,  as well as random values to resonator-TLS coupling term $\lambda^{(\alpha)}$ and the TLS $T^{(\alpha)}_1$ times, selected as in the previous section. The temperature  $T=0.09$ for all plots. 
\begin{figure}[htbp]
	\centering
		\includegraphics[height=2.2in]{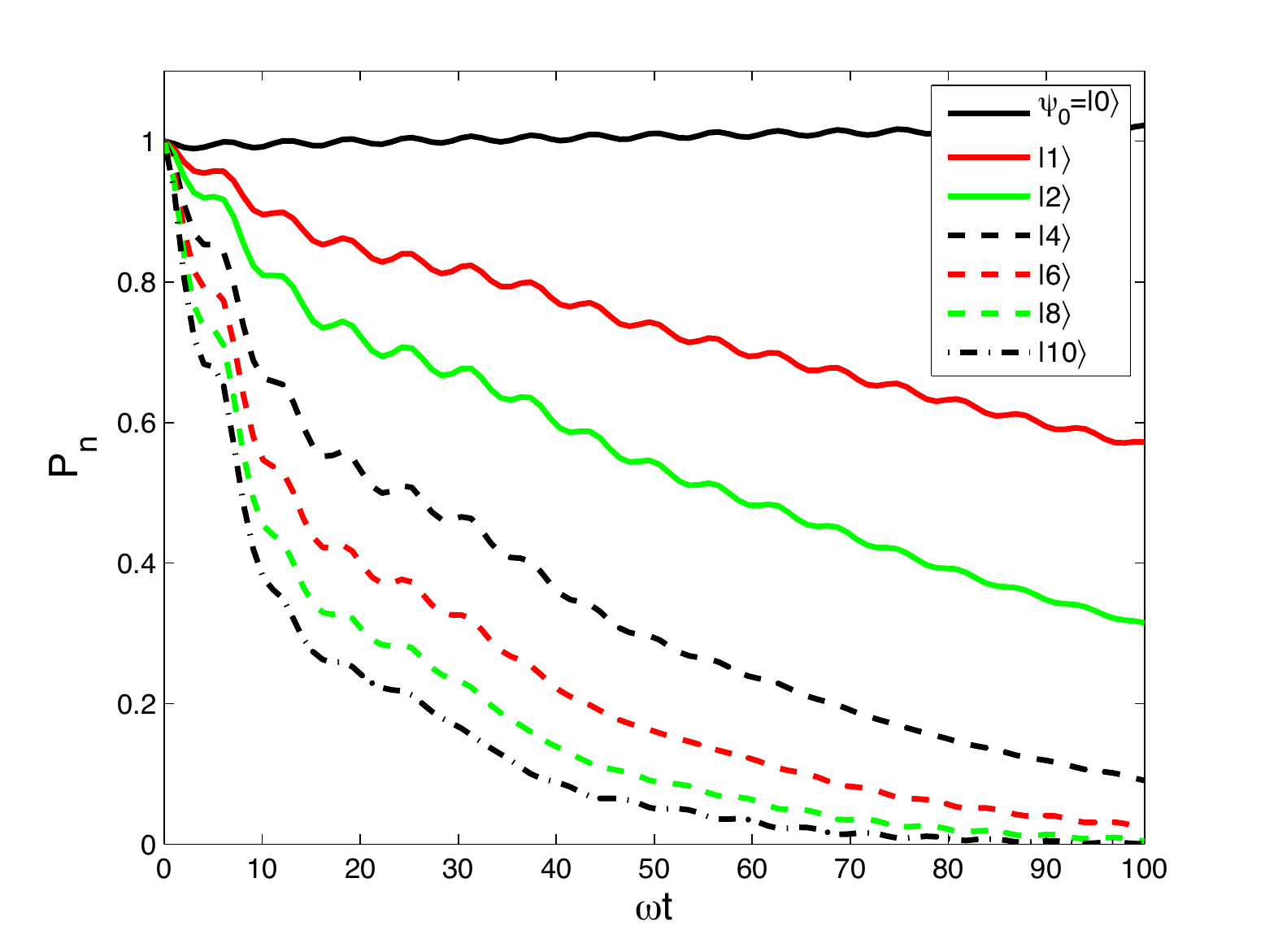}
		\includegraphics[height=2.2in]{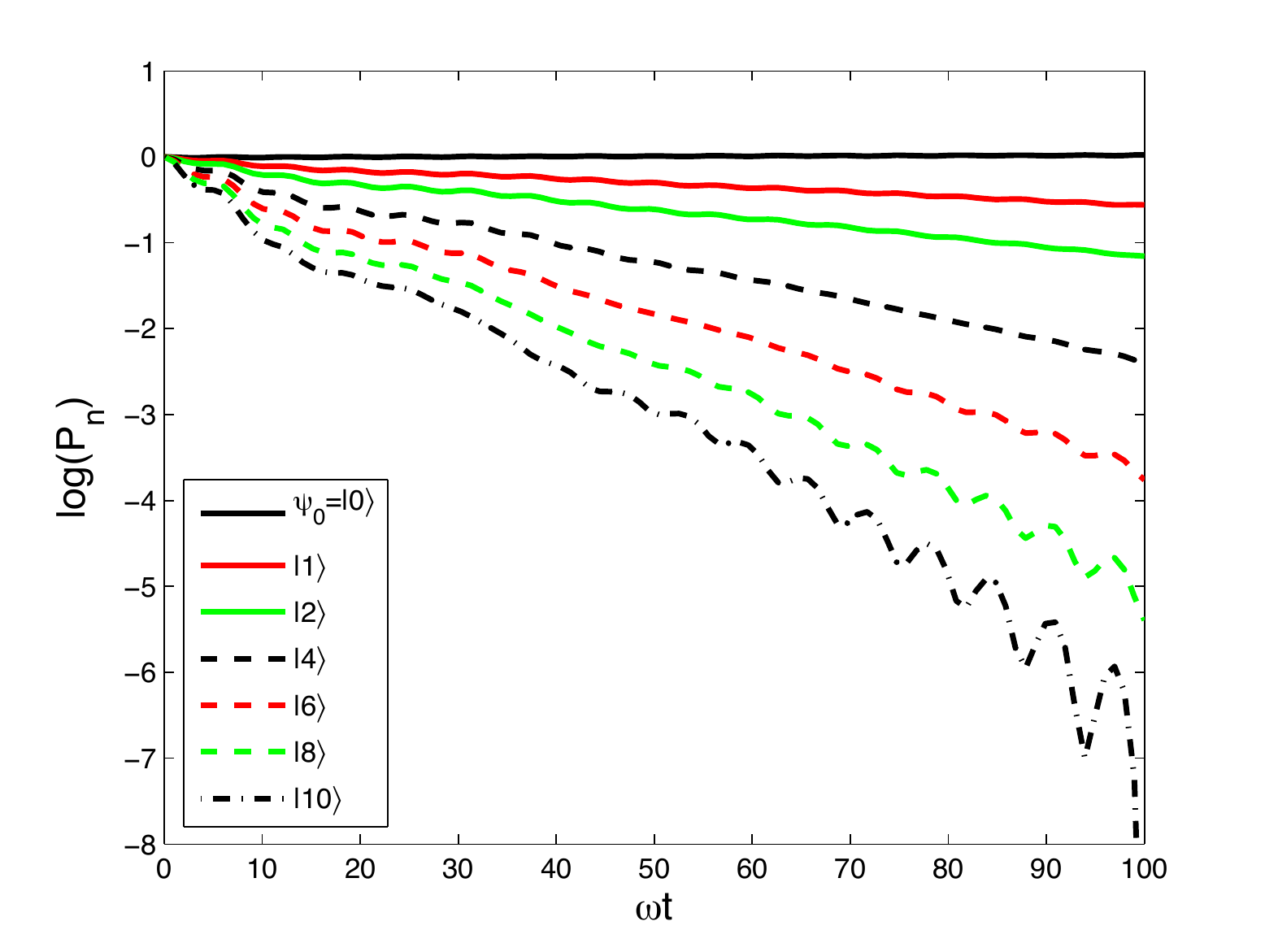}
	\caption{(Color online) Left: $P_n$ vs $\omega t$ for a resonator coupled to six non-interacting TLS's.  Right: Log of $P_n$ vs $\omega t$.  For all curves $T_1=10$.}
	\label{P_n6spins_no_int}
\end{figure}
We first consider the decay of a Fock state as a function of time for a resonator coupled to six damped spins.  From the log plot in Fig.~\ref{P_n6spins_no_int}, we see that  the natural log of the number state probability $P_n$ decays approximately linearly with time.  The oscillations at long times for the higher energy states are numerical artifacts arising from the exponentially small $P_n$ values.  We can apply a linear fit to the log plot to determine the $n$-dependence of the decay rate.  As for the resonator coupled to three TLS's, we find that the resonator's normalized decay rate scales with initial Fock state number similarly to that of an Ohmic bath, i.e. with slope $\approx -1$ (see Fig.~\ref{n_depend_6_spins_both}).  
\begin{figure}[htbp]
	\centering
		\includegraphics[height=3in]{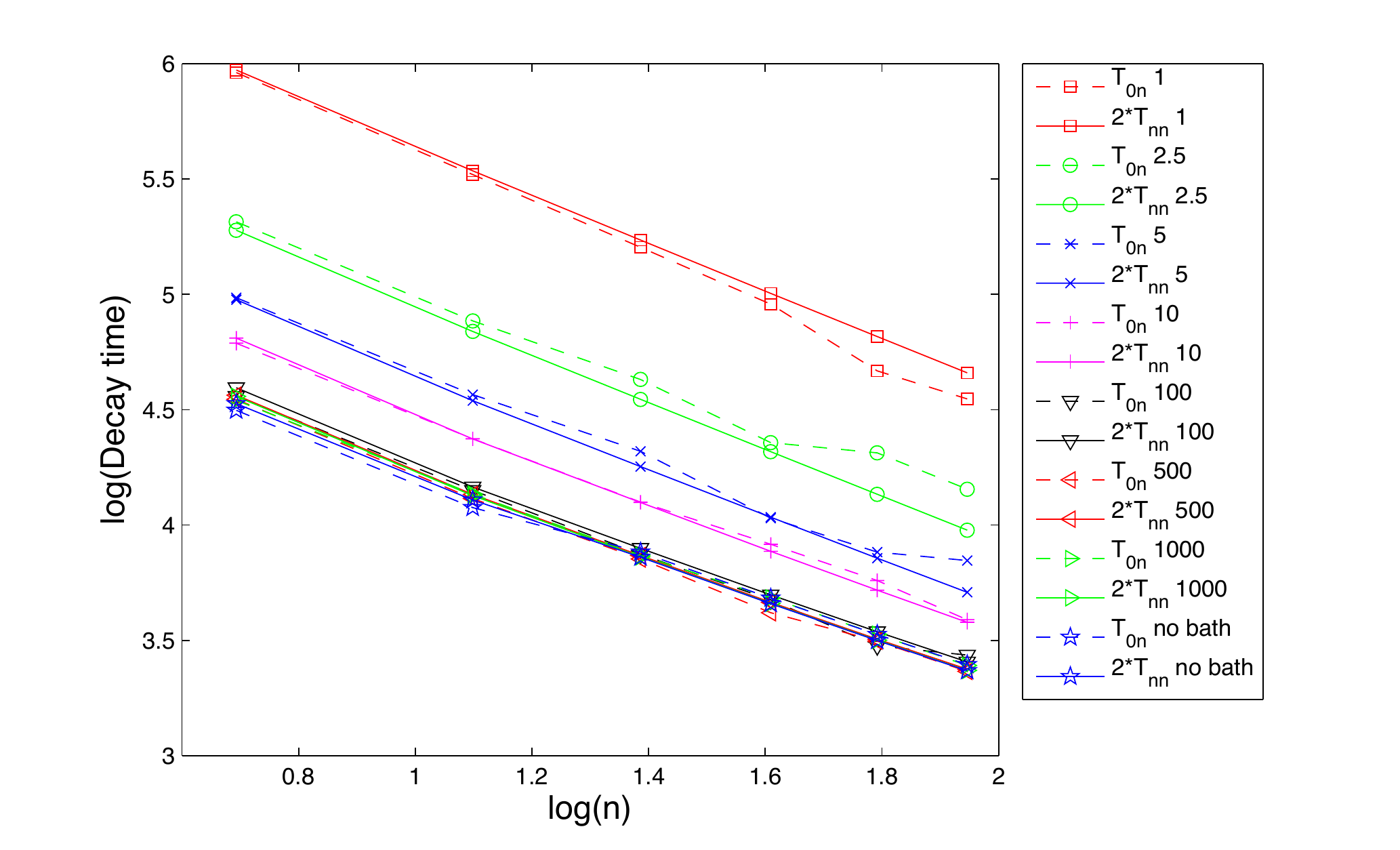}
	\caption{(Color online) $T_{0n}$ and $2T_{nn}$ for the initial superposition state $|\psi_0\rangle=1/\sqrt{2}(|0\rangle+|n\rangle)$ with a range of $T_1$ values.  For all curves the resonator is coupled to six non-interacting TLS's.  For the lowest curve (solid blue with star markers) the TLS's are not damped, and thus a $T_1$ time is not given.}
	\label{6spins_no_int}
\end{figure}

Next, we study the decay of a superposition of the ground state and the $n$th excited state.  Fig.~\ref{6spins_no_int} shows the log of $T_{0n}$ and $2T_{nn}$ vs the log of the initial $n$ characterizing the superposition state, for seven different values of the average TLS $T_1$ time.  We note that all of the curves have a slope $\approx-1$.   Similar to Fig.~\ref{3spins_no_int} for three TLS's, Fig.~\ref{6spins_no_int} shows little difference between $T_{0n}$ and $2T_{nn}$ for the different values of $T_1$; dephasing is negligible.

\begin{figure}[htbp]
	\centering
		\includegraphics[height=3in]{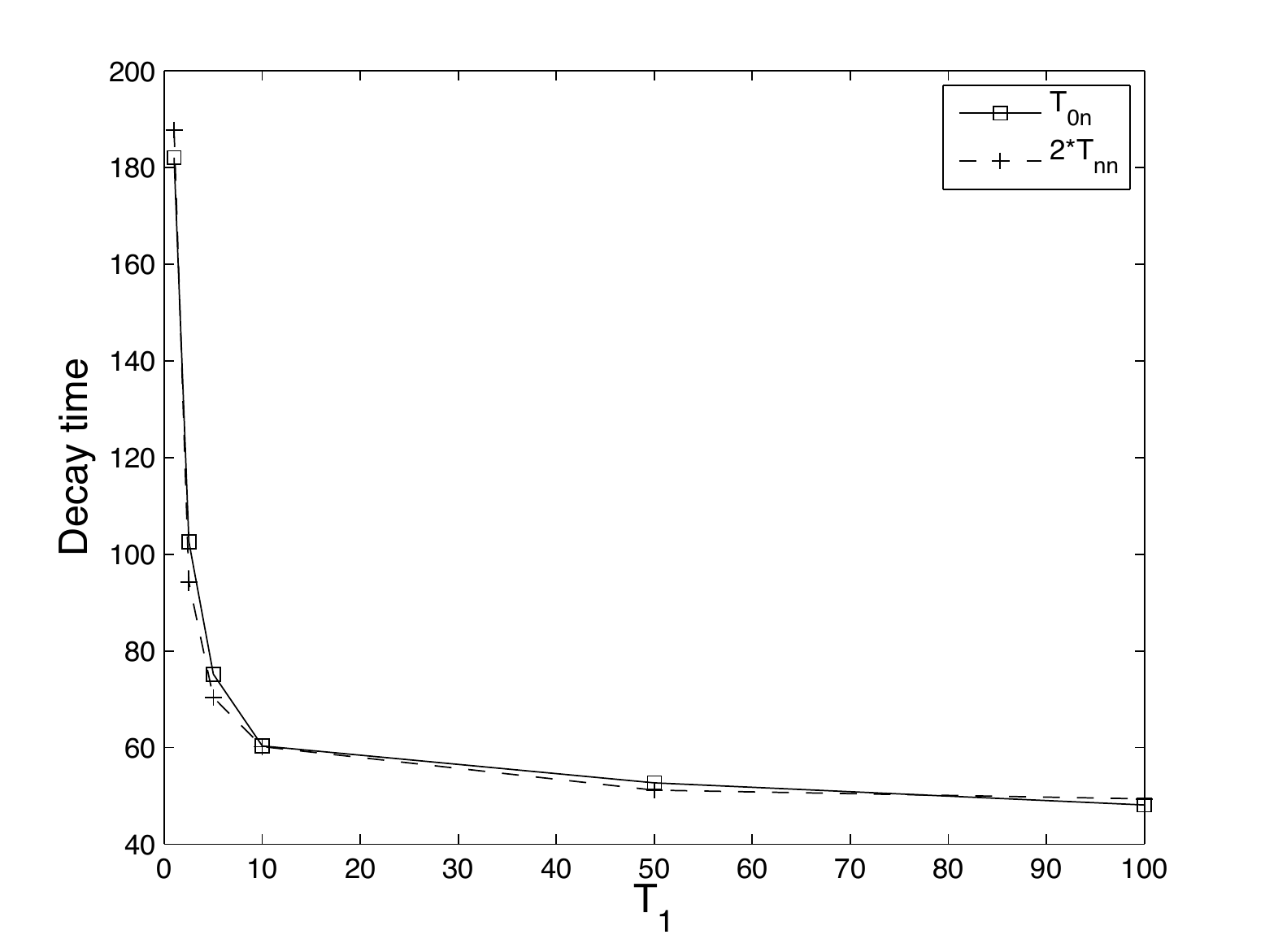}
	\caption{$2T_{nn}$ and $T_{0n}$ vs $T_1$ for the $n=4$ superposition state.}
	\label{T_16spins_no_int}
\end{figure}

In Fig.~\ref{T_16spins_no_int} we show the $T_1$ dependence of the on- and off-diagonal decay times for the $n=4$ superposition state.  The plot shows the same strong dependence on $T_1$ as for the case of three TLS's (Fig.~\ref{T_13spins_no_int}).  As a reminder, $T_1$ is the time it takes for a TLS in its excited state to decay to its ground state.  Thus, we would expect that as we decreased $T_1$ the resonator states would damp more quickly, resulting in a shorter decay time. For six TLS's, however, we find that as we decrease $T_1$, the diagonal and off-diagonal terms of the density matrix decay more slowly.  This unexpected behavior suggests that the coupling between the TLS's and their individual baths is somehow obstructing a more efficient means of dissipation.  This is supported by the lowest curve in Fig.~\ref{6spins_no_int}, which is for a resonator coupled to six TLS's that are {\it not} coupled to their individual baths, and yet indicates the shortest oscillator Fock state decay time. In Sec.~\ref{sec:approx},  we show through an analytical approximation that this behavior can be partially explained by considering  the  TLS bath Lorentzian line width dependencies on $T_1$ .

\begin{figure}[htbp]
	\centering
		\includegraphics[height=3in]{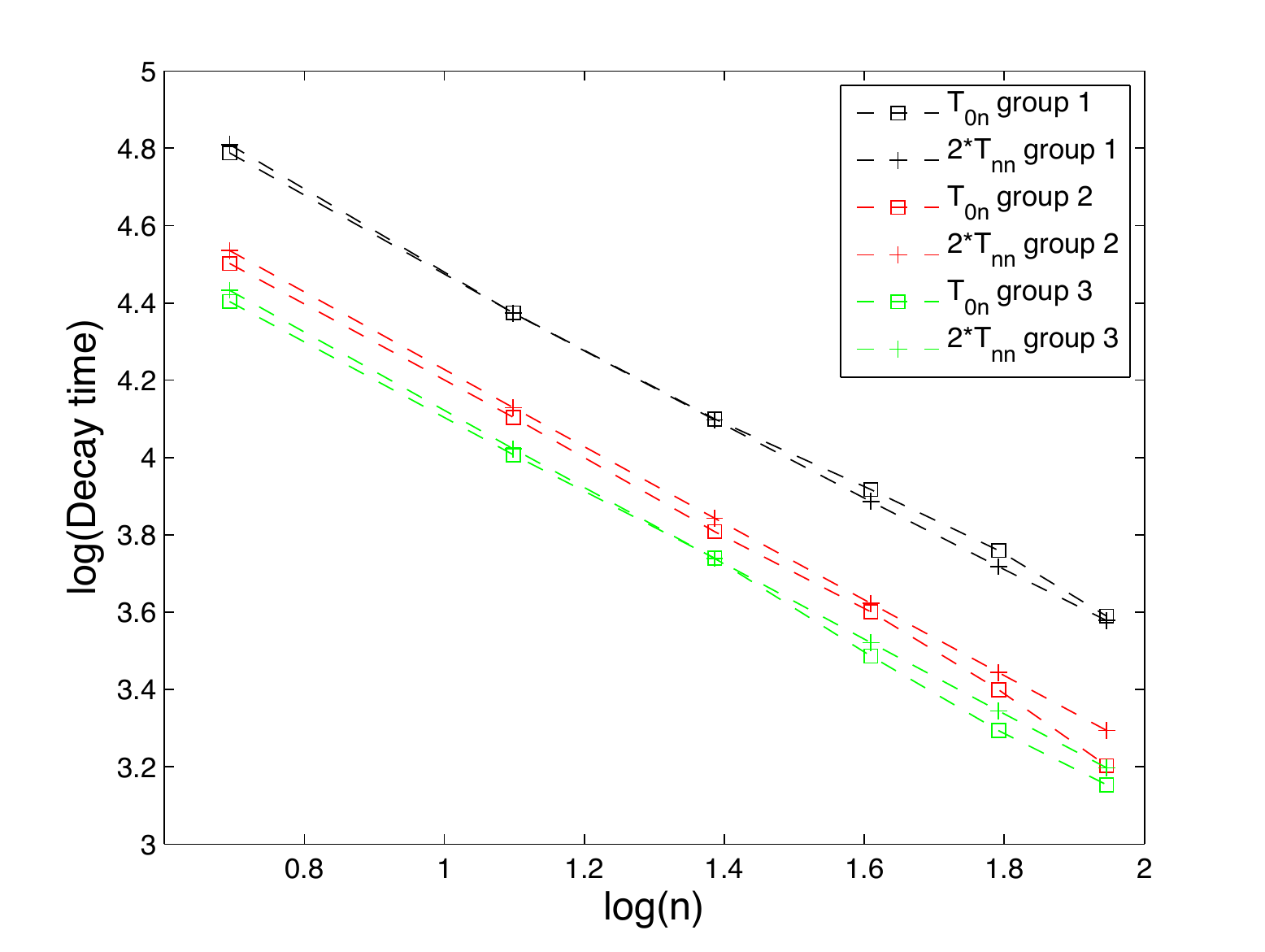}
	\caption{(Color online) $2T_{nn}$ and $T_{0n}$ vs $n$ for three different realizations of the TLS parameters.  For all curves $T_1=10$.}
	\label{3groups_6spins_no_int}
\end{figure}

Finally, as we did for three non-interacting TLS's, we now plot the decay of $T_{nn}$ and $T_{0n}$ for three different realizations of the TLS parameters.  Fig.~\ref{3groups_6spins_no_int} shows that the groups of six TLS's exhibit a higher degree of agreement than the three-TLS groups did, with uniform slopes $\approx-1$.  This is  a good indication that we have moved to a regime more akin to a dense TLS spectrum, with variations in the parameters of individual TLS's having less of an impact on the resonator. 

\subsection{\label{sec:approx}Analytical Approximation to Fock State Damping}

\begin{figure}[htbp]
	\centering
	\includegraphics[height=3in]{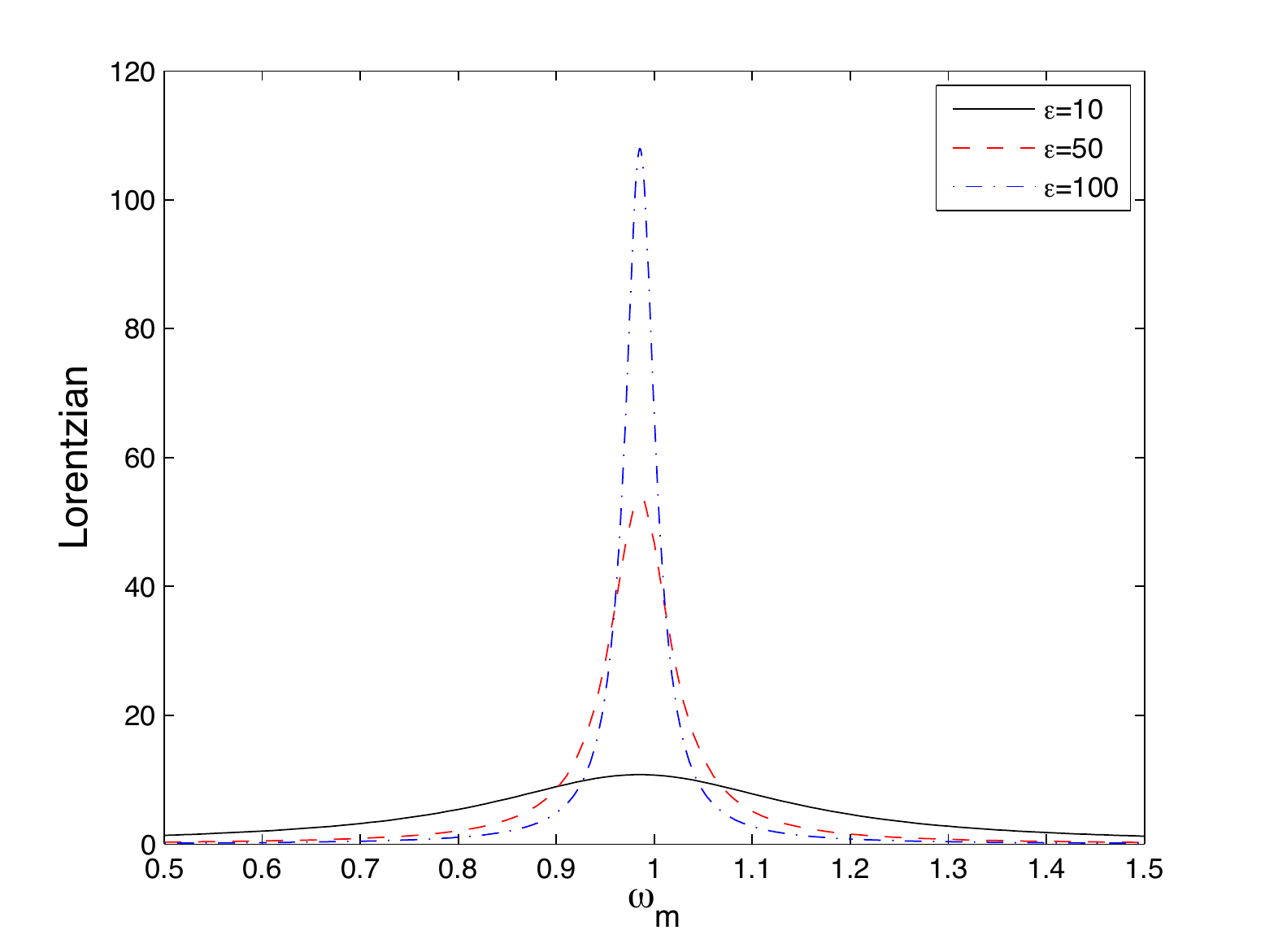} 
	\caption{(Color online) Lorentzian function vs $\omega_m$ for three different  $\epsilon$ values.  TLS parameters are $\Delta_0=0.6281$, $\Delta_b=0.7592$, and $T_1=0.6396$.}  
	\label{lorentzianfig}
\end{figure}

In this section we present an analysis of Fock state damping due to  TLS's.  We assume that the coupling between the mechanical resonator and the $N$ TLS's is sufficiently weak  that we can make a self-consistent Born approximation, where we expand perturbatively to second order in the  resonator-TLS couplings and trace over the TLS's to obtain the following resonator master equation: 
\begin{eqnarray}
\dot{\rho}_m (t)&=&-\frac{i}{\hbar}\left[H_m ,\rho_m (t)\right] \cr
&&-\frac{1}{\hbar^2}\int_0^t dt' \left\{\frac{1}{2}\langle\left\{B(t),B(t')\right\}\rangle\left[Y,\left[Y(t'-t),\rho_m (t)\right]\right]\right.\cr
&&+\left.\frac{1}{2}\langle\left[B(t),B(t')\right]\rangle\left[Y,\left\{Y(t'-t),\rho_m (t)\right\}\right]\right\},
\label{SCBAeq}
\end{eqnarray}
where $\rho_m$ and $H_m$ are the mechanical resonator density matrix and Hamiltonian, respectively, and 
\begin{equation}
B(t)=\sum_{\alpha=1}^N \lambda^{(\alpha)}  \sigma_z^{(\alpha)}(t),
\label{Beq}
\end{equation}
with $\lambda^{(\alpha)}$ the coupling between the oscillator and the $\alpha\mathrm{th}$ TLS.  Solving for the TLS-environment dynamics in the absence of the resonator, one can find the symmetric $\langle\{B(t),B(t')\}\rangle$ and antisymmetric $\langle[B(t),B(t')]\rangle$ correlation functions of the TLS bath.  Thus, in the above Born approximation, we neglect the influence of the resonator on the TLS dynamics.  More specifically, the approximation does not account for possible nonlinear, resonator amplitude-dependent saturation effects, or the possibility of coherent energy exchange between the resonator and the TLS's.  The importance of these effects depends on the relative coupling strengths between the mechanical resonator and the TLS's, and between the TLS's and their respective baths. 
Following the analysis in Ref.~[\onlinecite{shnirman05}], we have for the TLS bath correlation functions:
\begin{eqnarray}
\frac{1}{2}\langle\left\{B(t),B(t')\right\}\rangle&=&\sum_{\alpha=1}^N \left(\lambda^{(\alpha)}\right)^2 \left[\cos^2\theta^{(\alpha)} \left(1-\langle\sigma_z^{(\alpha)}\rangle^2\right) e^{-\Gamma_1^{(\alpha)} (t-t')}\right.\cr
&&\left. +\sin^2\theta^{(\alpha)}\cos\left[ E^{(\alpha)} (t-t')/\hbar\right]e^{-\Gamma_2^{(\alpha)} (t-t')} \right]
\label{symeq}
\end{eqnarray}  
and
\begin{eqnarray}
\frac{1}{2}\langle\left[B(t),B(t')\right]\rangle=-i\sum_{\alpha=1}^N \left(\lambda^{(\alpha)}\right)^2\sin^2\theta^{(\alpha)}\langle\sigma_z^{(\alpha)}\rangle\sin\left[ E^{(\alpha)} (t-t')/\hbar\right]e^{-\Gamma_2^{(\alpha)} (t-t')},
\label{antisymeq}
\end{eqnarray}  
 where $\langle\sigma_z^{(\alpha)}\rangle=\tanh\left(E^{(\alpha)}/(k_B T)\right)$, $\sin\theta^{(\alpha)}=\Delta_b^{(\alpha)}/E^{(\alpha)}$, and $\cos\theta^{(\alpha)}=\Delta_0^{(\alpha)}/E^{(\alpha)}$. The TLS dephasing rate is given in terms of the relaxation rate as
 \begin{equation}
 \Gamma_2^{(\alpha)}=\left[\frac{1}{2}+\left(\frac{\Delta_0^{(\alpha)}}{\Delta_b^{(\alpha)}}\right)^2\right]\Gamma_1^{(\alpha)},
 \label{dephaseeq}
 \end{equation}  
 where  $\Gamma_1^{(\alpha)}=T_1^{(\alpha)-1}$. We now substitute Eqs.~(\ref{symeq}) and (\ref{antisymeq}) into the mechanical resonator master equation~(\ref{SCBAeq}), and insert the free resonator (oscillator) dynamics solution 
 \begin{equation}
 Y(t'-t)=Y_{zp}\left(a e^{-i\omega_m(t'- t)} +a^\dagger e^{i\omega_m(t'- t)}\right).
 \end{equation}  
 We then make a rotating wave and a Markov approximation, and assume temperatures $k_BT\ll E^{(\alpha)}$ such that $\langle\sigma_z^{(\alpha)}\rangle\approx 1$.  We thus obtain the probability that the mechanical resonator is in the $n\mathrm{th}$ Fock state, $P_n =\langle n|\rho_m| n\rangle$: 
\begin{equation}
\frac{dP_n(t)}{dt}=-\gamma_{\mathrm{Fock}}\left[ nP_n(t)-(n+1)P_{n+1}(t)\right],
\label{probabilitydecayeq}
\end{equation}
where $\gamma_{\mathrm{Fock}} (\equiv T^{-1}_{11})$ gives the decay rate for an initial $n=1$ Fock state:
\begin{equation}
\gamma_{\mathrm{Fock}}=-\frac{1}{\hbar^2}\sum_{\alpha=1}^N \left(\lambda^{(\alpha)}\right)^2 \sin^2\theta^{(\alpha)} \frac{2\Gamma_2^{(\alpha)}}{\left(\Gamma_2^{(\alpha)}\right)^2+\left(E^{(\alpha)}/\hbar-\omega_m\right)^2}.
\label{fockdecayeq}
\end{equation} 
Eq.~(\ref{probabilitydecayeq}) shows that the decay rate for an initial $n$ Fock state scales with $n$, as we saw in the numerical simulations.  The dependence of the probability decay rate on $T_1$ comes from the TLS dephasing rate $\Gamma_2^{(\alpha)}$ dependence of the Lorentzian term.  From Eq.~(\ref{dephaseeq}), we see that $\Gamma_2^{(\alpha)}$ scales as $\Gamma_1^{(\alpha)}$.  We now consider the form of the given Lorentzian, subject to the rescaling $\epsilon T_1$:
\begin{equation}
\frac{2\Gamma_2/\epsilon}{\left(\Gamma_2/\epsilon\right)^2 +\left(E/\hbar-\omega_m\right)^2}.
\label{lorentzian}
\end{equation} 
Fig.~\ref{lorentzianfig} shows the Lorentzian factor as a function of $\omega_m$ for three different $\epsilon$ values.  
As we increase $\epsilon$ (i.e., increase the TLS damping time, $T_1$),  the Lorentzian factor  correspondingly increases, as long as $|E/\hbar -\omega_m|< \Gamma_2/\epsilon$, i.e., within the Lorentzian linewidth.  Physically, Eq.~(\ref{fockdecayeq}) indicates that for a mechanical resonator that is approximately resonant with a TLS, the longer the TLS decay time, the more rapidly it absorbs energy from the mechanical resonator, and hence the shorter the Fock state probability decay time.  However, as $\epsilon$ continues to increase, we eventually have that $|E/\hbar -\omega_m|> \Gamma_2/\epsilon$.  The TLS is no longer approximately resonant with the oscillator, and so the Lorentzian factor and thus the decay rate decreases. Fig.~{\ref{decaytime_vs_alpha} shows the dependence of the decay time on $\epsilon$ that follows from one of the distributions of TLS-oscillator coupling and parameter values used in the numerical simulations.  
\begin{figure}[htbp]
	\centering
		\includegraphics[height=3in]{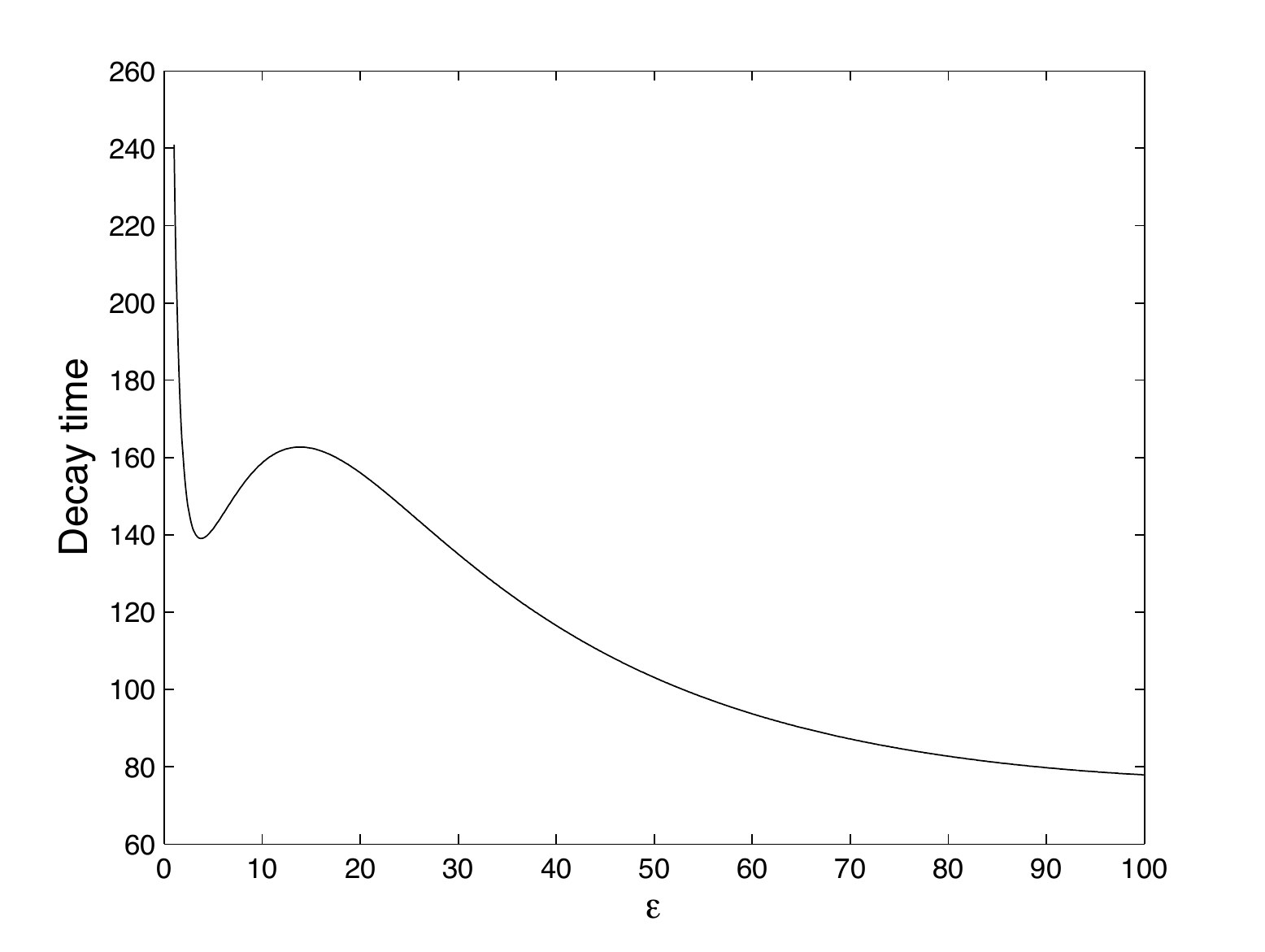} 
	\caption{Fock state decay time vs TLS $T_1$ scaling factor $\epsilon$ for the $n=1$ Fock state.}  
	\label{decaytime_vs_alpha}
\end{figure}
The intermediate dip is due to some of the TLS's going out of resonance. While the plot does not show quite the same monotonically decreasing decay time with increasing $\epsilon$ as found in the numerical simulation, it does give approximately the same overall decreasing trend. Differences are due to the breakdown of the Born-Markov approximation for treating the TLS subsystem as a bath.

\section{\label{sec:TLS-TLS}TLS-TLS Interactions}
\subsection{\label{hamiltonian}Derivation of Hamiltonian}
Experiments have shown that interactions between TLS's play an important role in dissipation and decoherence.\cite{arnold75,arnold78,graebner79}  In this section we derive the TLS-TLS interaction Hamiltonian.  We begin with the Hamiltonian for an elastic wave system interacting with TLS defects.  The Lagrangian for an elastic wave system is\cite{ashcroft76}
\begin{equation}
L_\mathrm{wave}=\frac{1}{2}\int_Vd^3r[\rho\dot{u}_i(\vec{r},t)\dot{u}_i(\vec{r},t)-c_{ijkl}\partial_iu_j(\vec{r},t)\partial_ku_l(\vec{r},t)],
\label{ElasticHameq}
\end{equation}
where $V$ is the system volume, $\rho$ is the mass density, $u_i(\vec{r},t), i=1,2,3$ is the $i$th component of the displacement vector field, and $c_{ijkl}$ is the elastic modulus tensor.  We use the Einstein summation convention.  The Hamiltonian is by definition
\begin{equation}
H_\mathrm{wave}=\dot{u}_i\frac{\partial L_\mathrm{wave}}{\partial\dot{u}_i}-L_\mathrm{wave},
\end{equation}
which, with Eq.~\ref{ElasticHameq}, gives
\begin{equation}
H_\mathrm{wave}=\int_Vd^3r\left[\frac{\rho}{2}\dot{u}_i\dot{u}_i+\frac{1}{2}c_{ijkl}\partial_iu_j\partial_ku_l\right].
\end{equation}
In addition to the non-interacting TLS Hamiltonian~(\ref{TLSHamiltonianeq}), we have the TLS-wave system interaction Hamiltonian
\begin{equation}
H_\mathrm{int}=-\sum_{\alpha=1}^N\left[\nu_{ij}^{(\alpha)}\epsilon_{ij}^{(\alpha)}\sigma_z^{(\alpha)}\right],
\end{equation}
where $\nu_{ij}^{(\alpha)}$ is the deformation potential tensor at the $\alpha$ TLS location $\vec{r}^{(\alpha)}$ and 
\begin{equation}
\epsilon_{ij}^{(\alpha)}=\frac{1}{2}[\partial_iu_j(\vec{r}^{(\alpha)},t)+\partial_ju_i(\vec{r}^{(\alpha)},t)]
\end{equation}
is the strain tensor at $\vec{r}^{(\alpha)}$.  Since $\nu_{ij}^{(\alpha)}=\nu_{ji}^{(\alpha)}$, we can rewrite the interaction Hamiltonian as 
\begin{equation}
H_\mathrm{int}=-\sum_{\alpha=1}^N\left[\nu_{ij}^{(\alpha)}\partial_iu_j(\vec{r}^{(\alpha)},t)\sigma_z^{(\alpha)}\right].
\end{equation}
The full Hamiltonian is now
\begin{eqnarray}
H &=& \int_Vd^3r\left[\frac{\rho}{2}\dot{u}_i\dot{u}_i+\frac{1}{2}c_{ijkl}\partial_iu_j\partial_ku_l\right]+\sum_{\alpha=1}^N \left[ \frac{1}{2}\Delta_0^{(\alpha)}\sigma_z^{(\alpha)}+\frac{1}{2}\Delta_b^{(\alpha)}\sigma_x^{(\alpha)}\right]\cr
&-&\sum_{\alpha=1}^N\left[\nu_{ij}^{(\alpha)}\partial_iu_j(\vec{r}^{(\alpha)},t)\sigma_z^{(\alpha)}\right].
\end{eqnarray}
For weak TLS-wave system interactions we can in principle start with this Hamiltonian and derive a master equation for the observed flexural wave mode of interest that interacts with the $N$ TLS's.  The rest of the elastic wave normal modes then form the TLS bath, as well as mediate the interactions between the TLS's.  Instead, we will adopt a less rigorous approach to derive the approximate form of the elastic wave-induced interaction between any pair of TLS's.
We assume that the timescale for the phonon mediated interaction between two TLS's is much shorter than their internal dynamics timescale; the two TLS's are therefore approximated as `frozen,' with each in a given spin state.  We take as our starting point the following Hamiltonian for the interaction between two TLS's without the non-interacting TLS part:     
\begin{equation}
H_{\mathrm{approx2TLS}}=\int_Vd^3r\left[\frac{\rho}{2}\dot{u}_i\dot{u}_i+\frac{1}{2}c_{ijkl}\partial_iu_j\partial_ku_l-\sum_{\alpha=1}^2\nu_{ij}^{(\alpha)}\partial_iu_j(\vec{r}^{(\alpha)},t)\sigma_z^{(\alpha)}\delta(\vec{r}-\vec{r}^{(\alpha)})\right].
\label{2TLSapproxHameq}
\end{equation}
Next, we express this approximate Hamiltonian operator at $t=0$ in terms of the normal mode, phonon creation and annihilation operators.  We define
\begin{equation}
u_i(\vec{r},0)=\sum_\beta\sqrt{\frac{\hbar}{2\rho\omega_\beta}}[a_\beta u_{\beta,i}(\vec{r})+a^\dagger_\beta u^*_{\beta,i}(\vec{r})]
\label{ueq}
\end{equation}
and
\begin{equation}
\dot{u}_i(\vec{r},0)=-i\sum_\beta\sqrt{\frac{\hbar\omega_\beta}{2\rho}}[a_\beta u_{\beta,i}(\vec{r})-a^\dagger_\beta u^*_{\beta,i}(\vec{r})],
\label{udoteq}
\end{equation}
where $[a_\beta,a_{\beta'}^\dagger]=\delta_{\beta,\beta'}$, with $\beta$ labeling the normal mode.  The normal modes are solutions to 
\begin{equation}
c_{ijkl}\partial_j\partial_ku_{\beta,l}=-\rho\omega_\beta^2u_{\beta,i}.
\label{normalmodeeq}
\end{equation}
Substituting Eqs.~(\ref{ueq}) and (\ref{udoteq}) into Eq.~(\ref{2TLSapproxHameq}) and using Eq.~(\ref{normalmodeeq}) and the orthonormality and completeness relations
\begin{equation}
\int_Vd^3ru_{\beta,i}(\vec{r})u^*_{\beta',i}(\vec{r})=\delta_{\beta,\beta'}
\end{equation}
and
\begin{equation}
\sum_\beta u_{\beta,i}(\vec{r})u^*_{\beta,j}(\vec{r}')=\delta_{ij}\delta(\vec{r}-\vec{r}'),
\label{completenesseq}
\end{equation}
respectively, we obtain
\begin{equation}
H_{\mathrm{approx2TLS}}=\sum_\beta\left[\frac{\hbar\omega_\beta}{2}(a_\beta a^\dagger_\beta+a^\dagger_\beta a_\beta)+f_\beta a_\beta+f^*_\beta a^\dagger_\beta\right],
\label{approx2TLSHam2eq}
\end{equation}
where the function $f_\beta$ is defined as
\begin{equation}
f_\beta=\sqrt{\frac{\hbar}{2\rho\omega_\beta}}\int_Vd^3ru_{\beta,i}(\vec{r})\nu_{ij}[\partial_j\delta(\vec{r}-\vec{r}^{(1)})\sigma_z^{(1)}+\partial_j\delta(\vec{r}-\vec{r}^{(2)})\sigma_z^{(2)}].
\label{fbetaeq}
\end{equation}
We now redefine the creation/annihilation operators so as to get rid of the linear operator terms in 
Eq.~(\ref{approx2TLSHam2eq}).  We make the substitution $\hat{a}_\beta=\hat{b}_\beta+c_\beta$, where we have included hats here to emphasize that $\hat{a}$ and $\hat{b}$ are operators, whereas $c$ is a commuting number.  We obtain (dropping the hats)
\begin{eqnarray}
H_\mathrm{approx2TLS} & = & \sum_\beta\left[\frac{\hbar\omega_\beta}{2}(b_\beta b^\dagger_\beta+b^\dagger_\beta b_\beta) + \hbar\omega_\beta(c^*_\beta b_\beta+c_\beta b^\dagger_\beta)\right.\nonumber\\
& &\left.+f_\beta b_\beta+f^*_\beta b^\dagger_\beta+c_\beta f_\beta+c^*_\beta f^*_\beta+\hbar\omega_\beta c_\beta c^*_\beta\right].
\label{approx2TLSHam3eq}
\end{eqnarray}
Defining $c_\beta=-f^*_\beta/\hbar\omega$ and $c^*_\beta=-f_\beta/\hbar\omega$, we see that the linear operator terms in Eq.~(\ref{approx2TLSHam3eq}) cancel out and we have
\begin{equation}
H_\mathrm{approx2TLS} = \sum_\beta\left[\frac{\hbar\omega_\beta}{2}(b_\beta b^\dagger_\beta+b^\dagger_\beta b_\beta)-\frac{f_\beta f^*_\beta}{\hbar\omega}\right].
\label{approx2TLSHam4eq}
\end{equation}
The TLS-TLS interaction term we are seeking is contained within the quadratic $f$-term in Eq.~(\ref{approx2TLSHam4eq}). Substituting in Eq.~(\ref{fbetaeq}) and simplifying, we obtain for the TLS-TLS interaction:
\begin{equation}
H_\mathrm{TLS-TLS}=-\sigma_z^{(1)}\sigma_z^{(2)}\nu_{ik}\nu_{jl}\frac{1}{\rho}\sum_\beta\frac{1}{\omega^2_\beta}\partial_ku_{\beta,i}(\vec{r}^{(1)})\partial_lu^*_{\beta,j}(\vec{r}^{(2)}), 
\label{TLS-TLSHameq}
\end{equation}
where we have neglected TLS self-interaction terms.  

The strength of the interaction between the two TLS's will depend on the nature of the elastic medium in which the TLS's are embedded, as expressed by the mode sum in Eq.~(\ref{TLS-TLSHameq}).
Let us now try to come up with a simple semiquantitative approximation to the mode sum part in the interaction term (\ref{TLS-TLSHameq}) using dimensional analysis.  From the completeness relation (\ref{completenesseq}), the displacement mode function $u_{\beta,i}$ has the dimensions $L^{-3/2}$ in terms of some to-be-determined length scale $L$. The mode frequency depends on the speed of sound $v$, and so scales as $\omega_\beta\sim v/L$.  Thus, the overall length dimension for the mode sum in Eq.~(\ref{TLS-TLSHameq}) is $L^{-3}$.  The relevant length scale, however, depends on the geometry of the embedding elastic medium.  For a bulk, three-dimensional (3D) medium where the two TLS's are far from any of the medium boundaries, the appropriate length scale must be the separation $r_{12}$ between the two TLS's.  Thus, for a 3D medium we have
\begin{equation}
H^\mathrm{3D}_\mathrm{2TLS}\sim\sigma_z^{(1)}\sigma_z^{(2)}\frac{\nu^2}{\rho v^2}\frac{1}{r^3_{12}},
\label{TLS-TLSHam3Deq}
\end{equation}
where we have neglected the anisotropy of the deformation potential.  For a membrane-like elastic medium, where the separation between the two TLS's is large compared to the membrane thickness $d$, we must lose one of the $r_{12}$ factors in (\ref{TLS-TLSHam3Deq}), to be replaced by $d$.  Thus, for an effectively 2D medium, we have
\begin{equation}
H^\mathrm{2D}_\mathrm{2TLS}\sim\sigma_z^{(1)}\sigma_z^{(2)}\frac{\nu^2}{\rho v^2 d}\frac{1}{r^2_{12}}.
\end{equation}
Finally, for a wire-like elastic medium where the separation between the two TLS's is large compared to the wire's crossectional dimensions $d$ and $w$, we must lose two of the $r_{12}$ factors in (\ref{TLS-TLSHam3Deq}).  Thus, for an effectively 1D medium, we have:
\begin{equation}
H^\mathrm{1D}_\mathrm{2TLS}\sim\sigma_z^{(1)}\sigma_z^{(2)}\frac{\nu^2}{\rho v^2 d w}\frac{1}{r_{12}}.
\end{equation}
Note that, as the dimensions of the elastic structure are reduced, the TLS-TLS interaction becomes longer ranged.  In particular, for a wire-like structure, the reduced volume and hence reduced number of TLS's will in part be compensated by a longer ranged interaction.

\subsection{\label{sec:3int}Three Interacting TLS's}

We now include TLS-TLS interactions.  We group all variables in Eq.~(\ref{TLS-TLSHameq}) except the sigma operators into a single variable, $\zeta^{(\alpha\beta)}$, simplifying the TLS-TLS interaction Hamiltonian to
\begin{equation}
H_\mathrm{TLS-TLS}=-\sigma_z^{(\alpha)}\sigma_z^{(\beta)}\zeta^{(\alpha\beta)}.
\end{equation}
For the plots in this section and the next, we choose a value for $\zeta$ and then generate a random $\zeta^{(\alpha\beta)}$ within $\pm50\%$ of this value for each pair of TLS's.  Unless otherwise specified, the values are centered around $\zeta=0.1/6$.  We plot $P_n$ vs $\omega t$ for a resonator coupled to three interacting TLS's.

Fig.~\ref{P_n3spins_int} shows that $P_n$ decays exponentially as a function of time, and we again apply a linear fit to the log plot to extract a decay rate.  In Fig.~\ref{n_depend_3spins_both} we plot the normalized decay rate for a resonator coupled to three non-interacting TLS's (solid), to three interacting TLS's (dash), and to an Ohmic bath (dot-dash).  The dot-dashed and dashed curves are practically indistinguishable, suggesting that the addition of TLS-TLS interactions allows the three TLS's to absorb energy like an Ohmic bath, even for higher-$n$ Fock states.
\begin{figure}[htbp]
	\centering
		\includegraphics[height=2.2in]{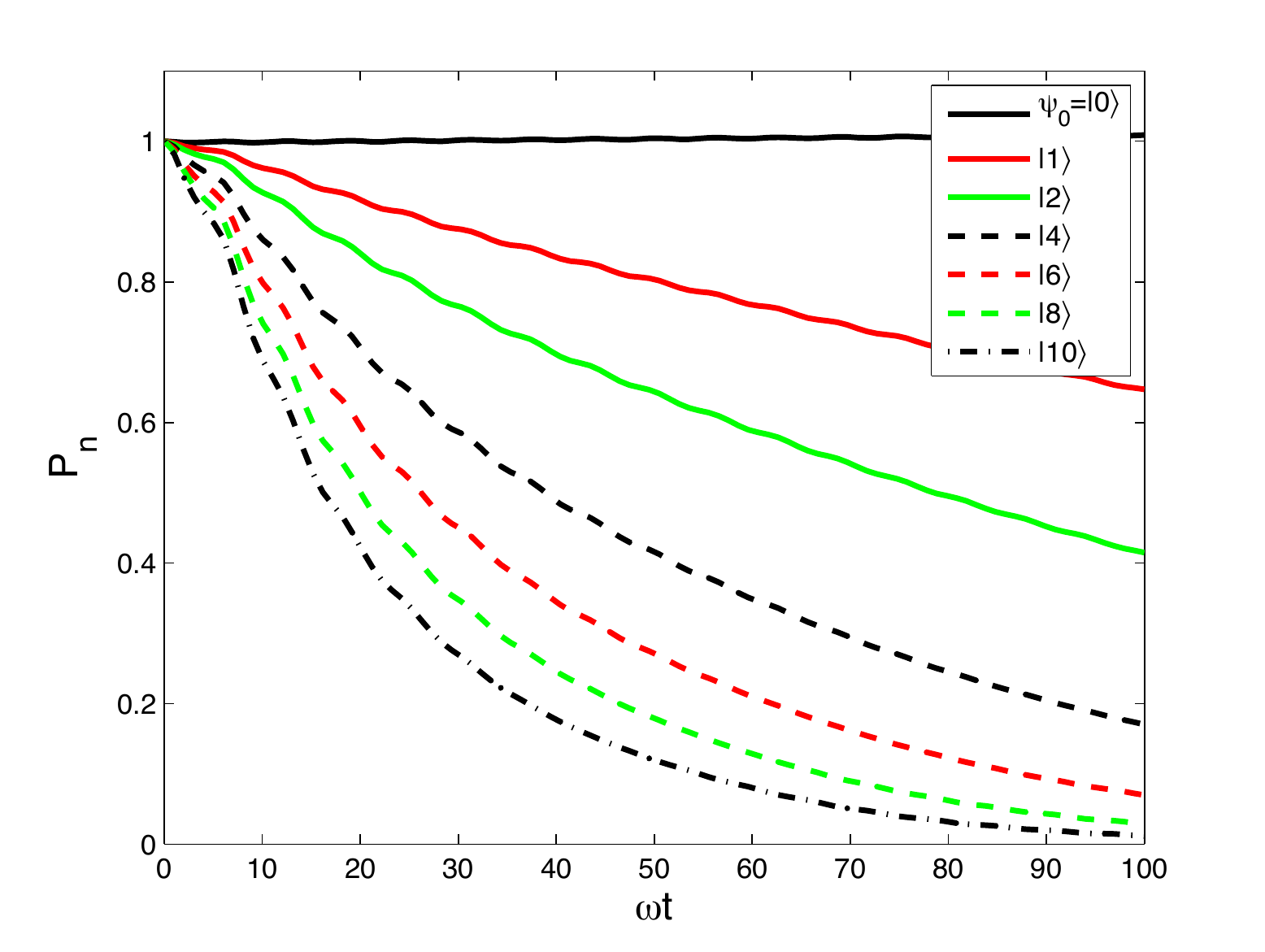}
		\includegraphics[height=2.2in]{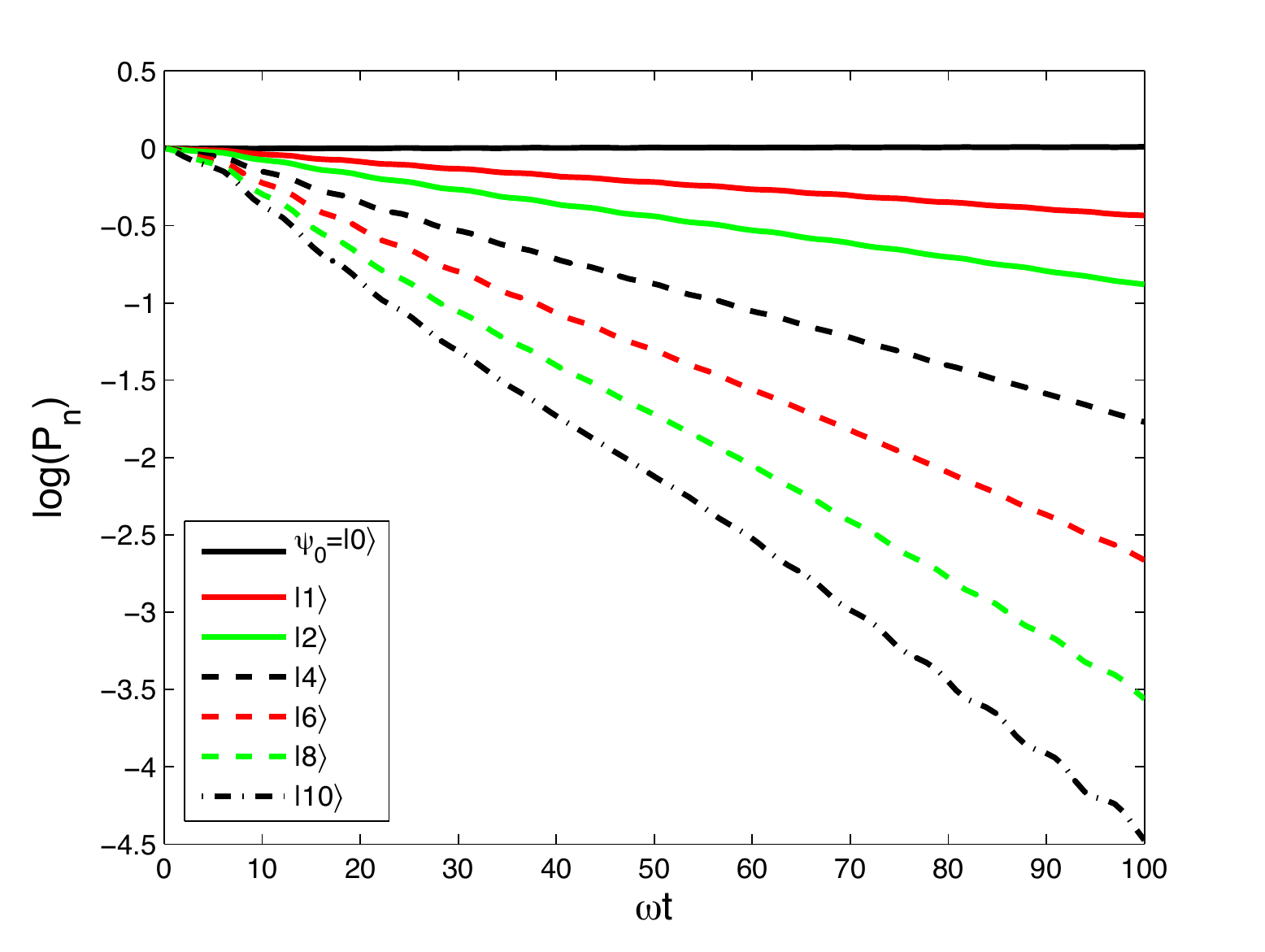}
	\caption{(Color online) Left: $P_n$ vs $\omega t$ for a resonator coupled to three interacting TLS's.  Right: Log 	of $P_n$ vs $\omega t$.  For all curves $T_1=10$.}
	\label{P_n3spins_int}
\end{figure}
\begin{figure}[htbp]
	\centering
		\includegraphics[height=3in]{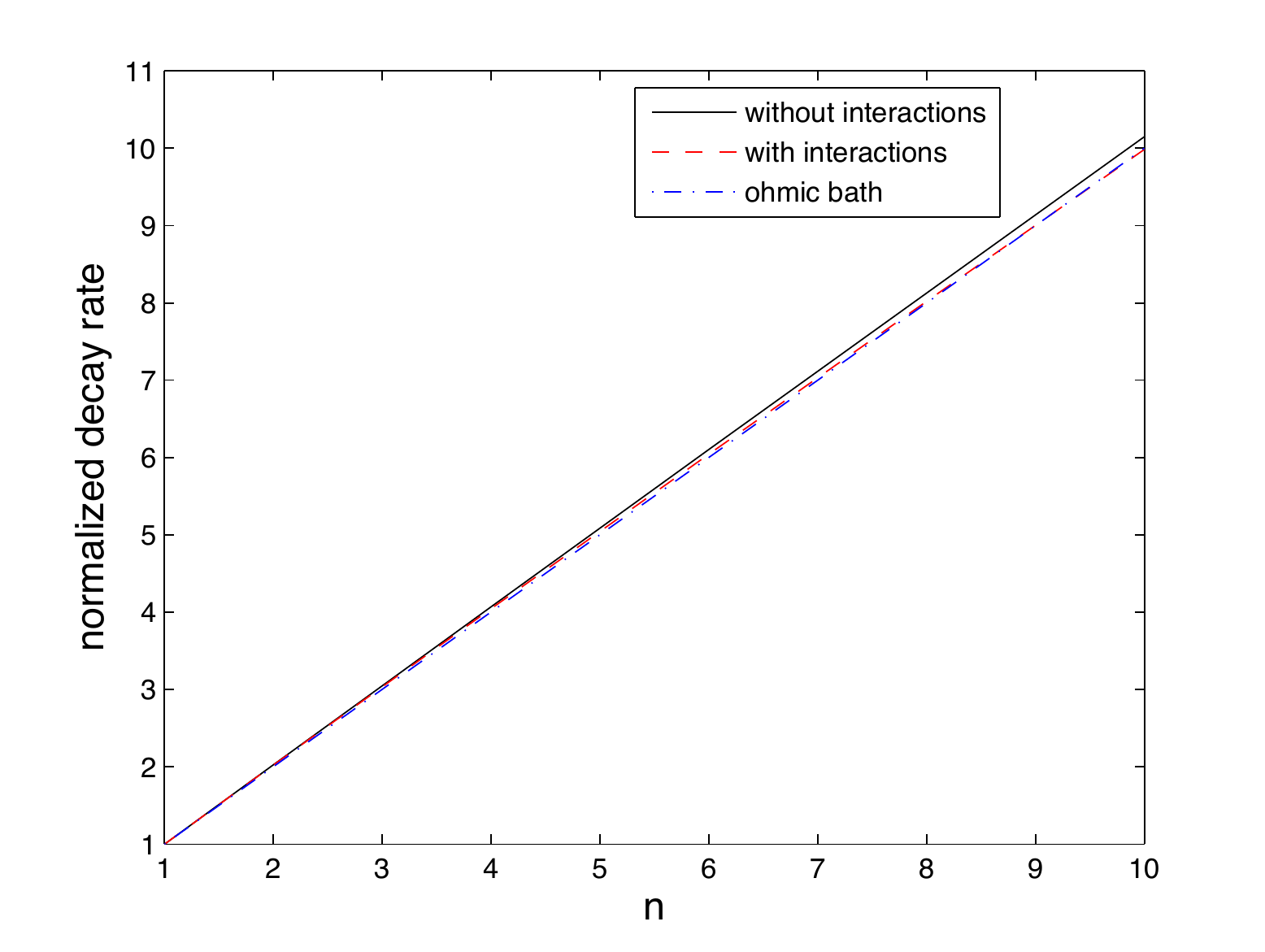}
	\caption{(Color online) Normalized decay rate vs $n$ for single Fock states. The resonator is coupled to three non-interacting TLS's (solid), three interacting TLS's (dash), and to an Ohmic bath (dot-dash).  For all curves $T_1=10$.}
	\label{n_depend_3spins_both}
\end{figure}

We now investigate the decay of a superposition state.  Fig.~\ref{3spins_int} shows the on- and off-diagonal decay times $T_{0n}$ and $2T_{nn}$ as a function of $n$ for a range of TLS $T_1$ values.  As in the case of three non-interacting TLS's, the curves have slopes $\approx-1$ and also $T_{0n}\approx 2T_{nn}$; dephasing is negligible.
\begin{figure}[htbp]
	\centering
		\includegraphics[height=3in]{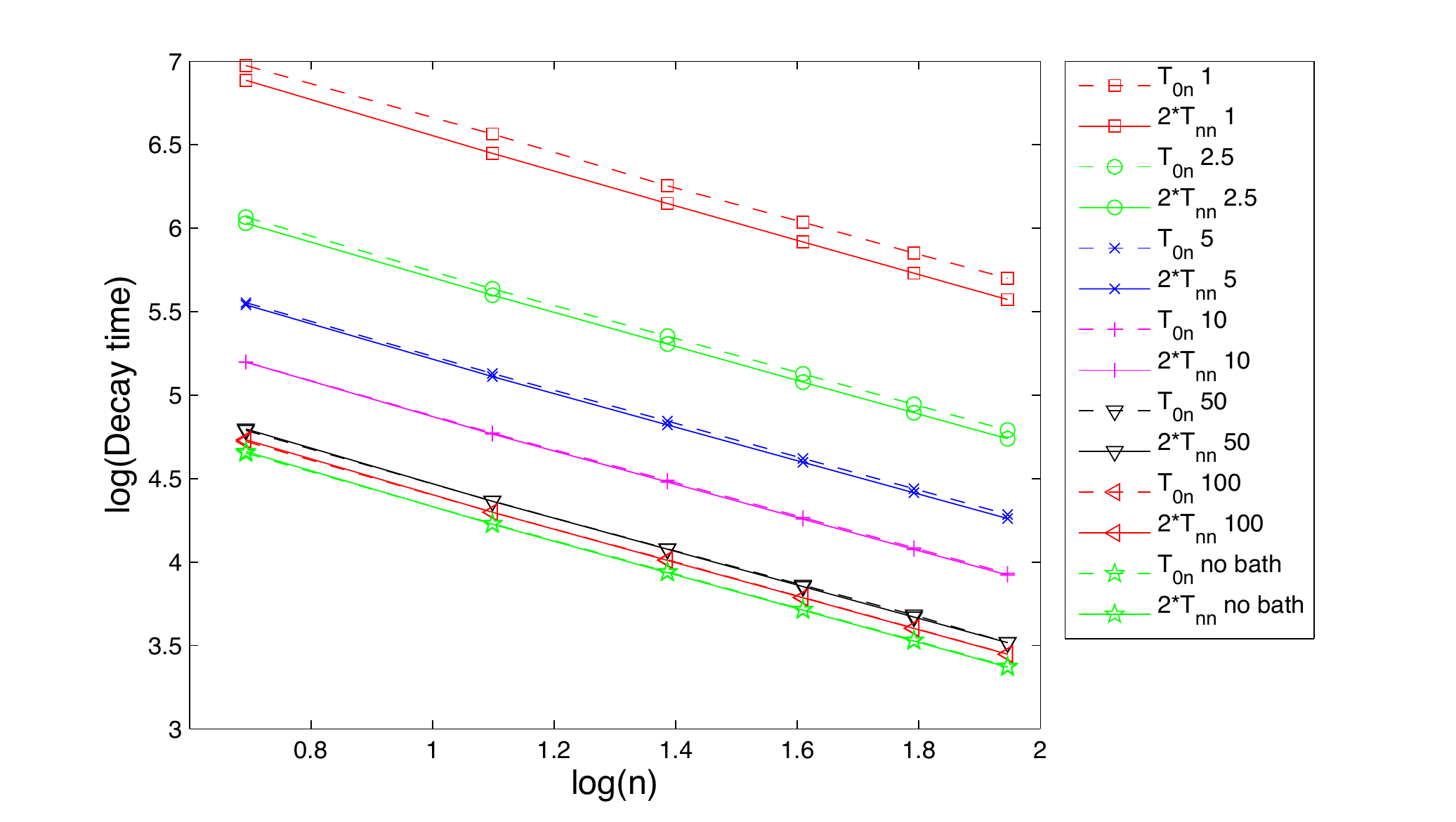}
	\caption{(Color online) $T_{0n}$ and $T_{nn}$ for a superposition state $|\psi_0\rangle=1/\sqrt{2}(|0\rangle+|n\rangle)$ for a range of $T_1$ values.  The resonator is coupled to three interacting TLS's.  For the final curve the TLS's are not damped, and thus a $T_1$ time is not given.}
	\label{3spins_int}
\end{figure} 

Fig.~\ref{T_13spins_both} shows the $T_1$ dependence of the on- and off-diagonal decay times for a resonator coupled to three non-interacting (black) and three interacting (gray) TLS's.  The decay times are reduced for the resonator coupled to interacting TLS's, with the same unexpected $T_1$ dependence as noted in the previous section.  
\begin{figure}[htbp]
	\centering
		\includegraphics[height=3in]{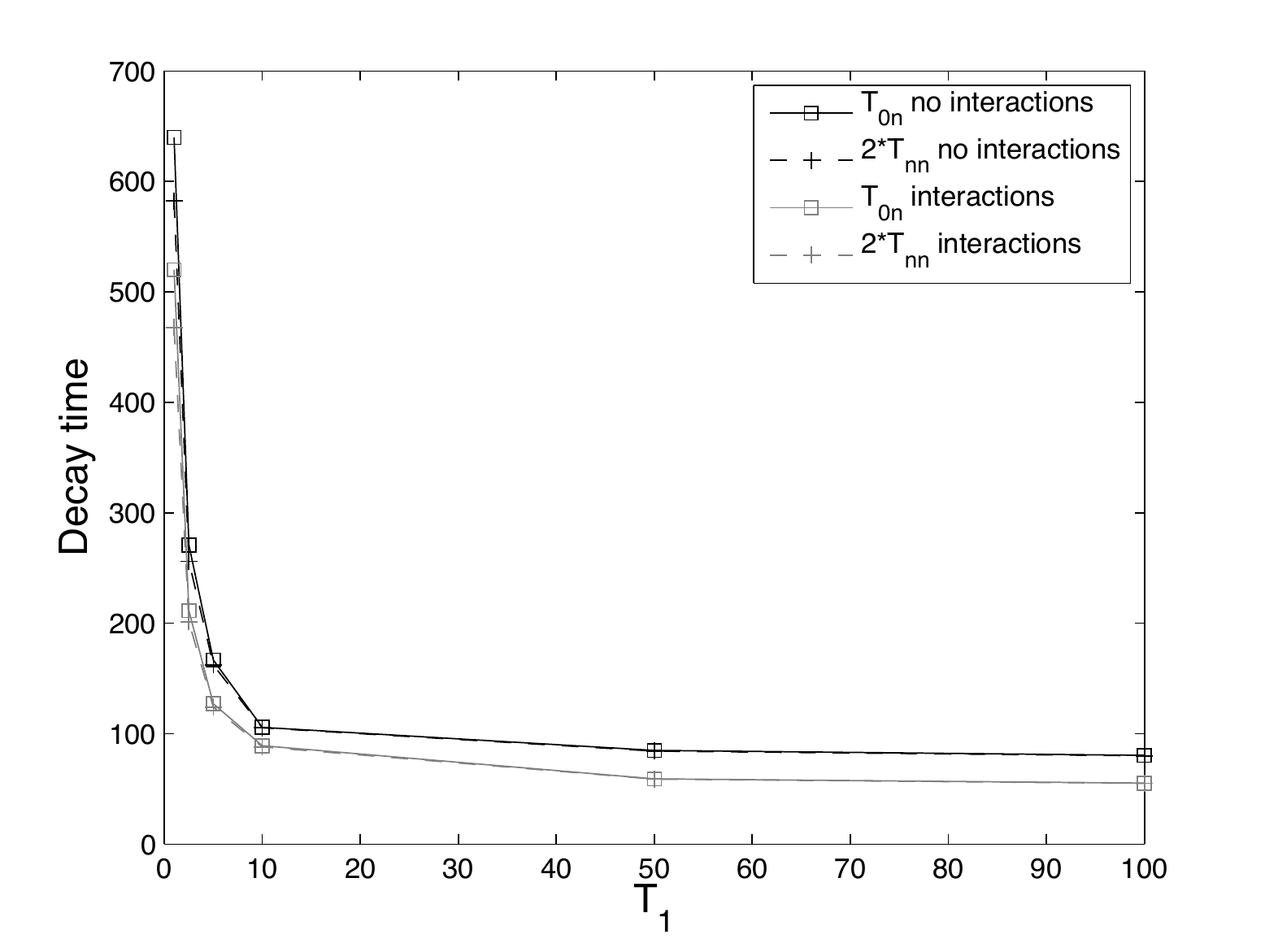}
	\caption{$T_{nn}$ and $T_{0n}$ vs $T_1$ for the $n=4$ superposition state. The resonator is coupled to three non-interacting (black) and three interacting (gray) TLS's.}
	\label{T_13spins_both}
\end{figure}

\subsection{\label{sec:6int}Six Interacting  TLS's}

We now couple the resonator to six interacting TLS's.  Fig.~\ref{P_n6spins_int} shows the number state probability as a function of time for a resonator coupled to six damped, interacting TLS's.  The shape of the curves is similar to that for six non-interacting TLS's, with the log plot appearing approximately linear. Again, the oscillations appearing in the larger $n$ curves at long times are numerical artifacts due to the exponentially small decay probabilities.   In Fig.~\ref{n_depend_6_spins_both} we plot the decay rate as a function of $n$ for six TLS's with (dash) and without (solid) TLS-TLS interactions, and for a resonator coupled only to an Ohmic bath (dot-dash).  We note that the decay is similar for the two cases with slope close to one.  
\begin{figure}[htbp]
	\centering
		\includegraphics[height=2.2in]{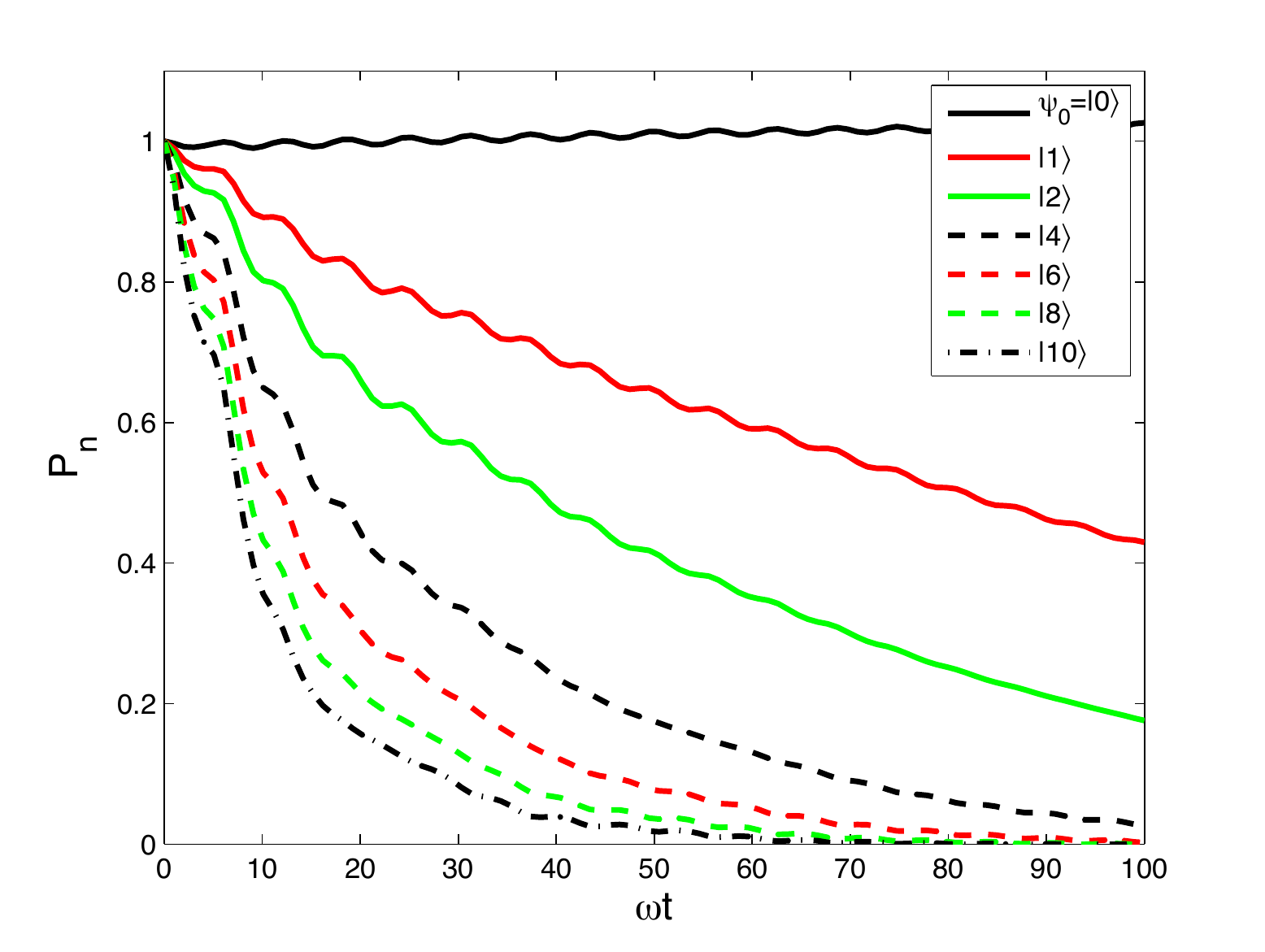}
		\includegraphics[height=2.2in]{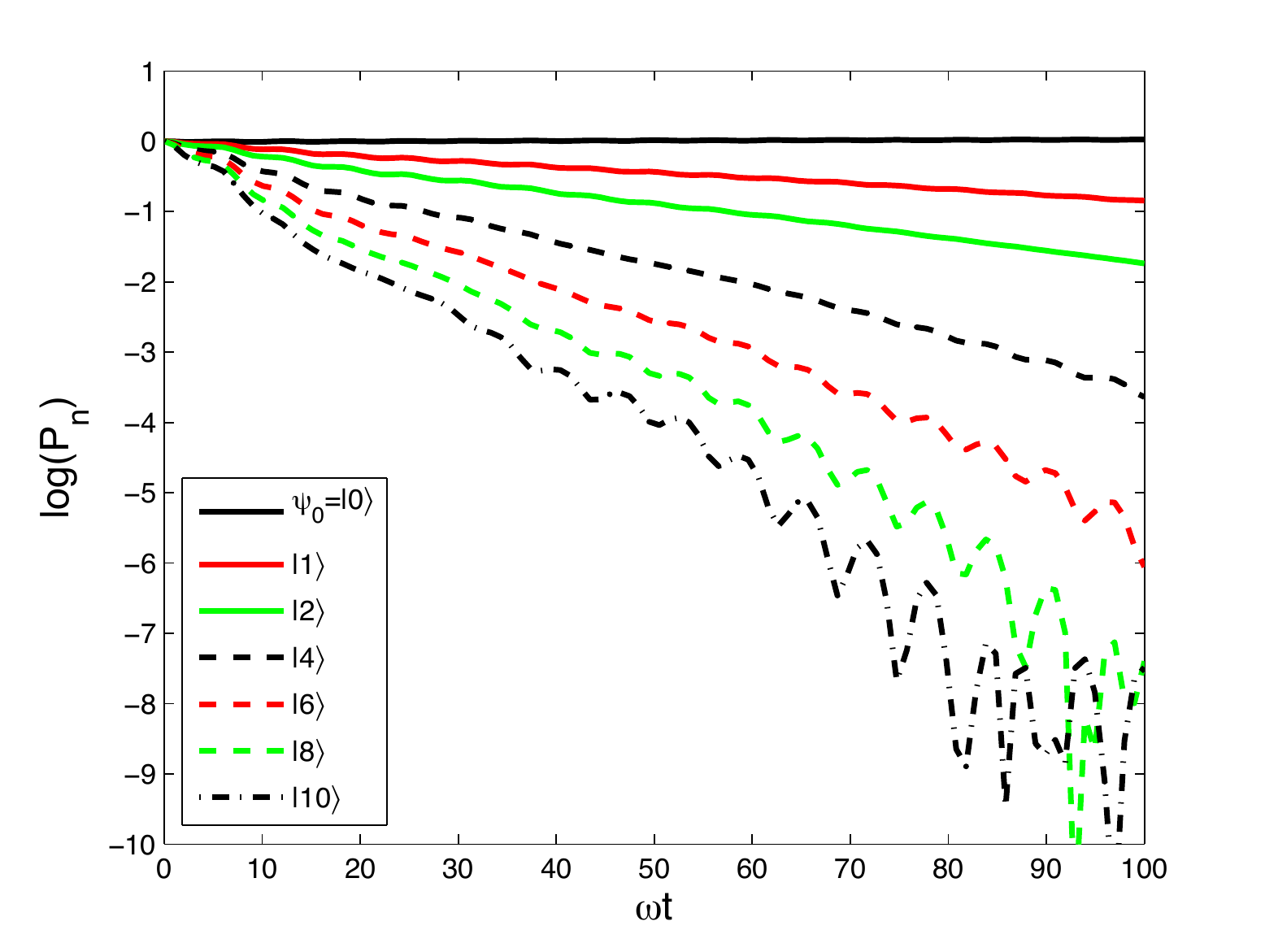}
	\caption{(Color online) Left: $P_n$ vs $\omega t$ for a resonator coupled to six interacting TLS's.  Right: Log of $P_n$ vs $\omega t$.  For all curves $T_1=10$.}
	\label{P_n6spins_int}
\end{figure}

Next we study the $T_1$ dependence of the on- and off-diagonal terms of the density matrix for a superposition of Fock states, as we did in  Fig.~\ref{T_13spins_both}.  Fig.~\ref{T_16spins_both} shows $T_{nn}$ and $T_{0n}$ as a function of $T_1$ for the $n=4$ superposition state for a resonator coupled to six non-interacting (black) and six interacting (gray) TLS's.  The plot shows that in both cases a reduction of $T_1$ causes an increase in the decay time of the on- and off-diagonal terms.  As for the three interacting TLS case, the addition of TLS-TLS interactions decreases the decay times.
\begin{figure}[htbp]
	\centering
		\includegraphics[height=3in]{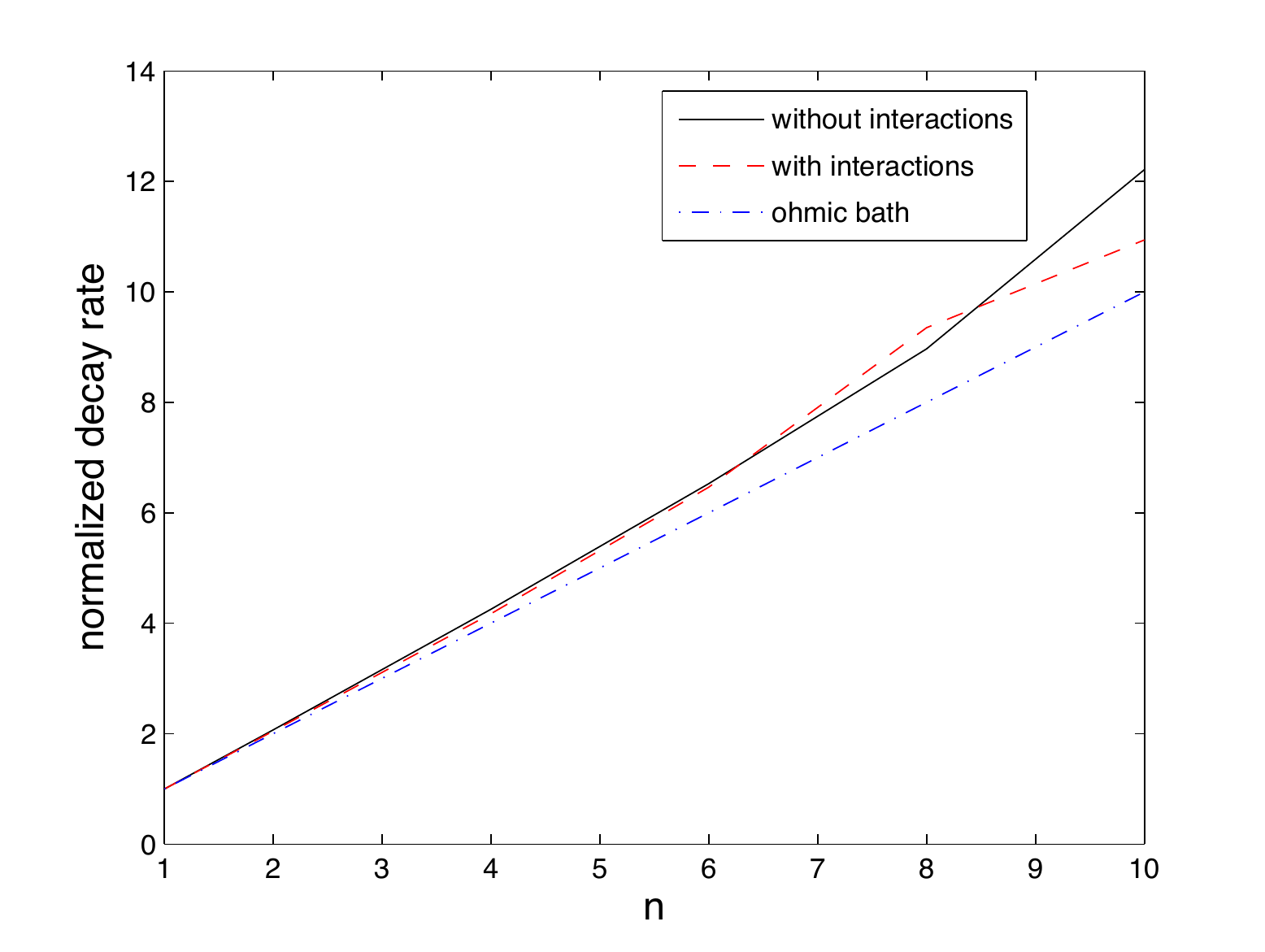}
	\caption{(Color online) Normalized decay rate vs $n$ for a single Fock state. The resonator is coupled to six non-interacting TLS's (solid), six interacting TLS's (dash), and to an Ohmic bath (dot-dash).  For all curves $T_1=10$.}
	\label{n_depend_6_spins_both}
\end{figure}
\begin{figure}[htbp]
	\centering
		\includegraphics[height=3in]{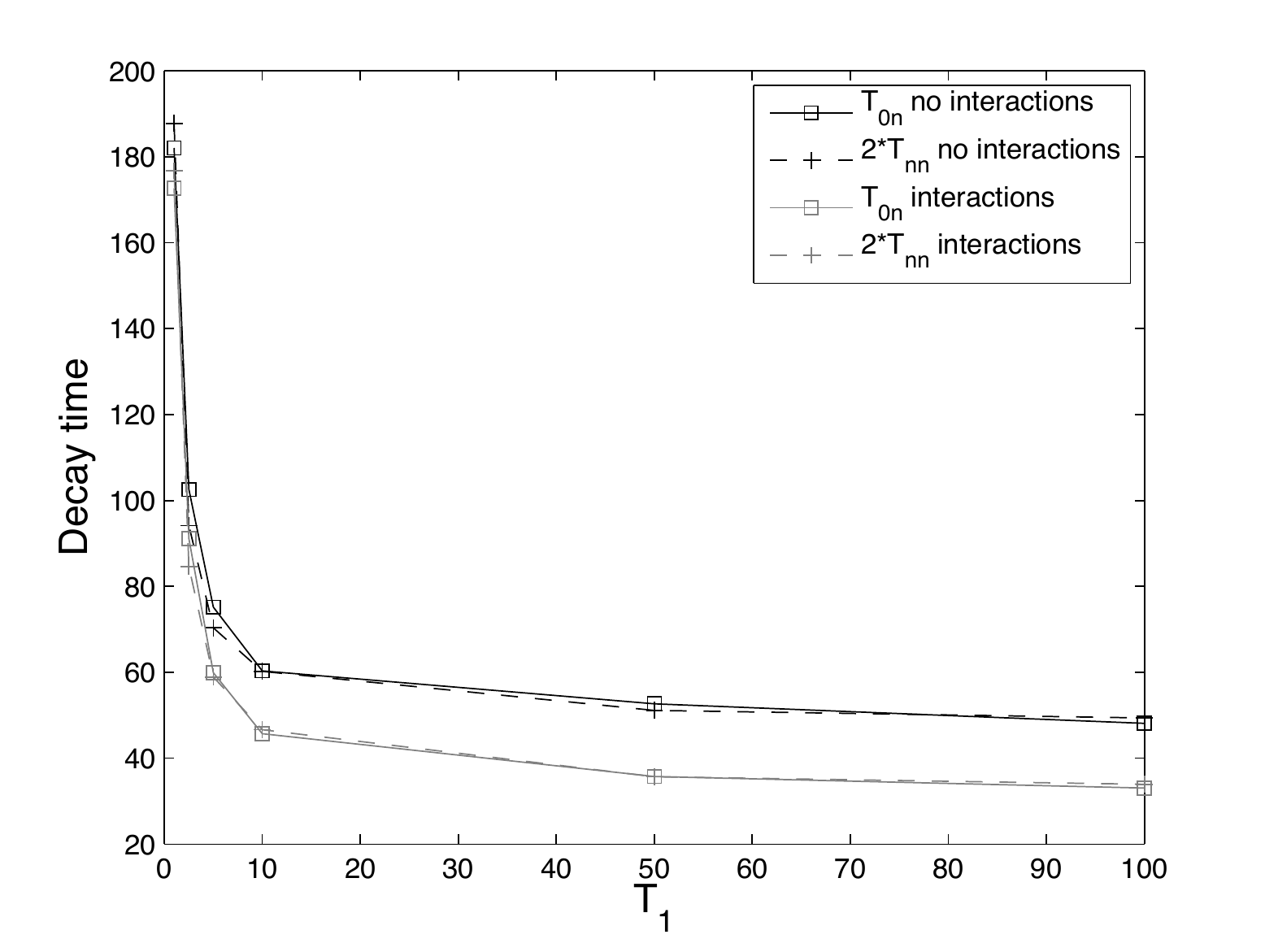}
	\caption{$T_{nn}$ and $T_{0n}$ vs $T_1$ for the $n=4$ superposition state. The resonator is coupled to six non-interacting (black) and six interacting (gray) TLS's.}
	\label{T_16spins_both}
\end{figure}

Lastly, we plot $T_{nn}$ and $T_{0n}$ as a function of the TLS-TLS coupling parameter $\zeta$.  Fig.~\ref{zeta6spins} shows the $\zeta$ dependence of the decay times for two different sets of random $\zeta^{(\alpha\beta)}$.  In both cases the decay time of the diagonal terms, $T_{nn}$, shows a linear dependence on the strength of the TLS-TLS coupling, with a slight variation in the slope for the two realizations.  The off-diagonal terms, particularly for the second group of random $\zeta^{(\alpha\beta)}$ values, decay less uniformly with respect to $\zeta$, but the overall behavior shows a clear dependence on $\zeta$, with stronger TLS-TLS coupling leading to faster decay of both the diagonal and off-diagonal terms of the density matrix. 
\begin{figure}[htbp]
	\centering
		\includegraphics[height=3in]{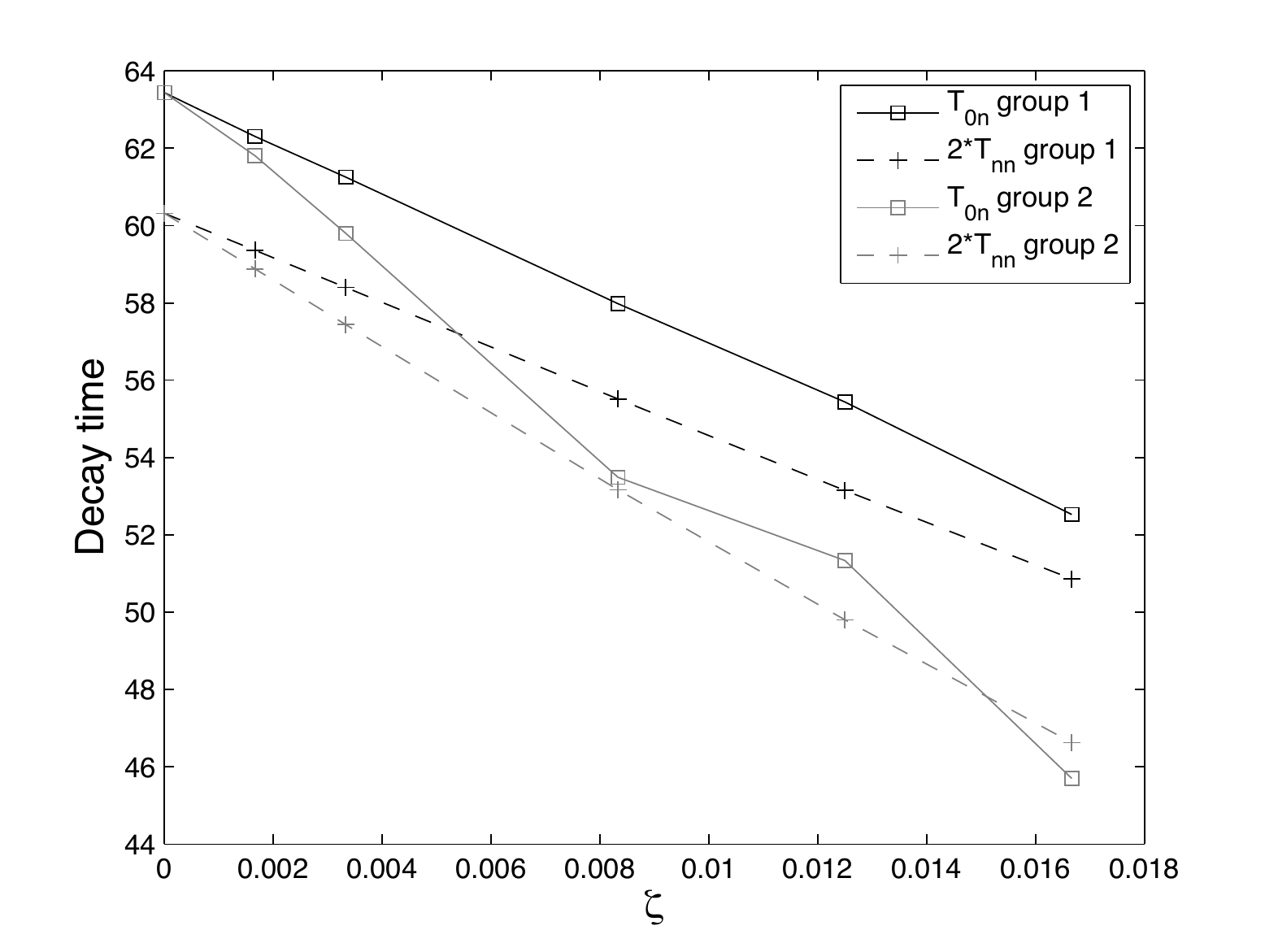}
	\caption{$T_{nn}$ and $T_{0n}$ vs the TLS-TLS interaction strength $\zeta$ for two different realizations of the random $\zeta^{(ij)}$ values.  The resonator is initially in the $n=4$ superposition state.  For all curves $T_1=10$.}
	\label{zeta6spins}
\end{figure}

\section{\label{sec:conclusion}conclusion}
In this work we have explored the effects of TLS's on damping and decoherence of Fock states in a mechanical resonator at low temperatures.  We began our investigation with a resonator coupled to a single TLS, and then increased the number to three, and then six TLS's.  For Fock states in a resonator coupled to a single, damped TLS, we observed amplitude-dependent damping, an indication of TLS saturation.  For a resonator coupled to three TLS's we found that the decay of Fock states and Fock state superpositions was similar to that due to an Ohmic bath, particularly when TLS-TLS interactions were included.  We noted that there was still some variation in the decay for different realizations of the random TLS variables, reflecting the fact that we were between a single TLS and a dense spectrum of TLS's.   For a resonator in the presence of six TLS's we found that the damping of a Fock state went approximately as $T_1/n$, as expected for an Ohmic bath.  Further, for the decay of a superposition of Fock states we found that the off-diagonal terms of the density matrix decayed twice as slowly as the diagonal terms, as expected for a resonator coupled to a bath of free oscillators.  While the on- and off-diagonal terms showed an unexpected dependence on $T_1$, the analysis in Sec.~\ref{sec:approx} suggested a possible explanation for their behavior in terms of $T_1$ dependences  of the TLS decay line widths.  

This work highlights the need for analytical approximations in order to understand the numerical results.  One possibility is the use of a polaron-like transformation to take into account the correlation between the TLS's and the resonator.\cite{tian10}  This may allow us to simplify by approximation the equations of the oscillator and damped TLS's in order to understand, for example, the qualitative saturation dependence.  Furthermore, the Ohmic bath-like dependence, i.e., decay rate proportional to $n$, for several near-resonant interacting TLS's suggests that appropriate analytical methods can help understand the relevant simulations.  Much work remains to be done to understand the quantum-classical correspondence, particularly now that experiments demonstrating mechanical systems in the quantum limit are becoming a reality. 

\section*{Acknowledgements}
We thank Susan Schwarz for her assistance in implementing the numerical simulations on Dartmouth's Discovery Cluster.This work was supported by the National Science Foundation under Grants No. DMR-0804477 and No. DMR-1104790.

\end{document}